\documentclass[aps,pre,superscriptaddress,floats,floatfix,twocolumn]{revtex4-1}

\usepackage{mathptmx}      
\usepackage[utf8]{inputenc}
\usepackage[T1]{fontenc}
\usepackage[german,english]{babel}
\usepackage{amsmath, amssymb}
\usepackage{graphicx, subfigure}
\usepackage{bm} 
\usepackage{textcomp}

\renewcommand{\vec}{\mathbf}  
\let\nablaOld\nabla
\renewcommand{\nabla}{\vec\nablaOld}  
\newcommand{\de}{\mathop{}\!\mathrm{d}}  
\DeclareMathOperator{\artanh}{Artanh}  

\begin{document}
\title{Local Lorentz force and ultrasound Doppler velocimetry in a vertical convection liquid metal flow}
\author{Till Zürner}
  \email[Corresponding author: ]{till.zuerner@tu-ilmenau.de}
  \affiliation{Technische Universität Ilmenau, Institut für Thermo- und Fluiddynamik, Postfach 100565, D-98684 Ilmenau, Germany}
\author{Tobias Vogt}
  \affiliation{Helmholtz-Zentrum Dresden -- Rossendorf, Institut für Fluiddynamik, Abteilung Magnetohydrodynamik, Bautzner Landstraße 400, 01328 Dresden, Germany}
\author{Christian Resagk}
  \affiliation{Technische Universität Ilmenau, Institut für Thermo- und Fluiddynamik, Postfach 100565, D-98684 Ilmenau, Germany}
\author{Sven Eckert}
  \affiliation{Helmholtz-Zentrum Dresden -- Rossendorf, Institut für Fluiddynamik, Abteilung Magnetohydrodynamik, Bautzner Landstraße 400, 01328 Dresden, Germany}
\author{Jörg Schumacher}
  \affiliation{Technische Universität Ilmenau, Institut für Thermo- und Fluiddynamik, Postfach 100565, D-98684 Ilmenau, Germany}
\date{November 6, 2017}
\begin{abstract}
We report velocity measurements in a vertical turbulent convection flow cell that is filled with the eutectic liquid metal alloy gallium-indium-tin by the use of local Lorentz force velocimetry (LLFV) and ultrasound Doppler velocimetry (UDV). We demonstrate the applicability of LLFV for a thermal convection flow and reproduce a linear dependence of the measured force in the range of micronewtons on the local flow velocity magnitude. Furthermore, the presented experiment is used to explore scaling laws of the global turbulent transport of heat and momentum in this low-Prandtl-number convection flow. Our results are found to be consistent with theoretical predictions and recent direct numerical simulations.
\end{abstract}
\maketitle
\section{Introduction}
\label{sec:introduction}
Despite numerous technological applications, such as in material processing \citep{Davidson2001,Asai2012,Shevchenko2013} or in liquid metal batteries \citep{Kelley2014}, convective flow phenomena in liquid metals are still much less well studied than in air or water \citep{Chilla2012}. The velocity measurement cannot rely on standard optical methods such as particle image velocimetry \citep{Adrian2011} or particle tracking and requires alternative methods. Ultrasound Doppler velocimetry \citep{Takeda1986,Brito2001,Eckert2002} and X-ray radiography \citep{Boden2008} are two non-invasive methods for opaque liquid metal fluids in laboratory experiments.\par
However, the high electrical conductivity of liquid metals  with values larger than $10^6$\,S/m opens the possibility of inductive measurement methods. These include invasive techniques such as Vives probes \citep{Ricou1982}, where a small permanent magnet is inserted into the liquid and the potential drop across the magnet surface due to the flow is measured. A similar, but non-invasive technique is electrical potential velocimetry (EPV) \citep{Baker2017}. Here multiple electrodes are embedded in the wall of the container and a global magnetic field is applied. The measured potential differences between electrodes give the 2D velocity field close to the wall. Since it is contact-based, EPV still has to cope with potentially aggressive fluids. The measurement of the induced magnetic field by a flow in an external magnetic field is used e.g. by contactless inductive flow tomography (CIFT), which is non-invasive and, as the name implies, contactless \citep{Wondrak2017}. An extensive list of further methods is given in \citet{Heinicke2013a}.\par
\begin{figure*}[t]
\includegraphics[height=.8\columnwidth]{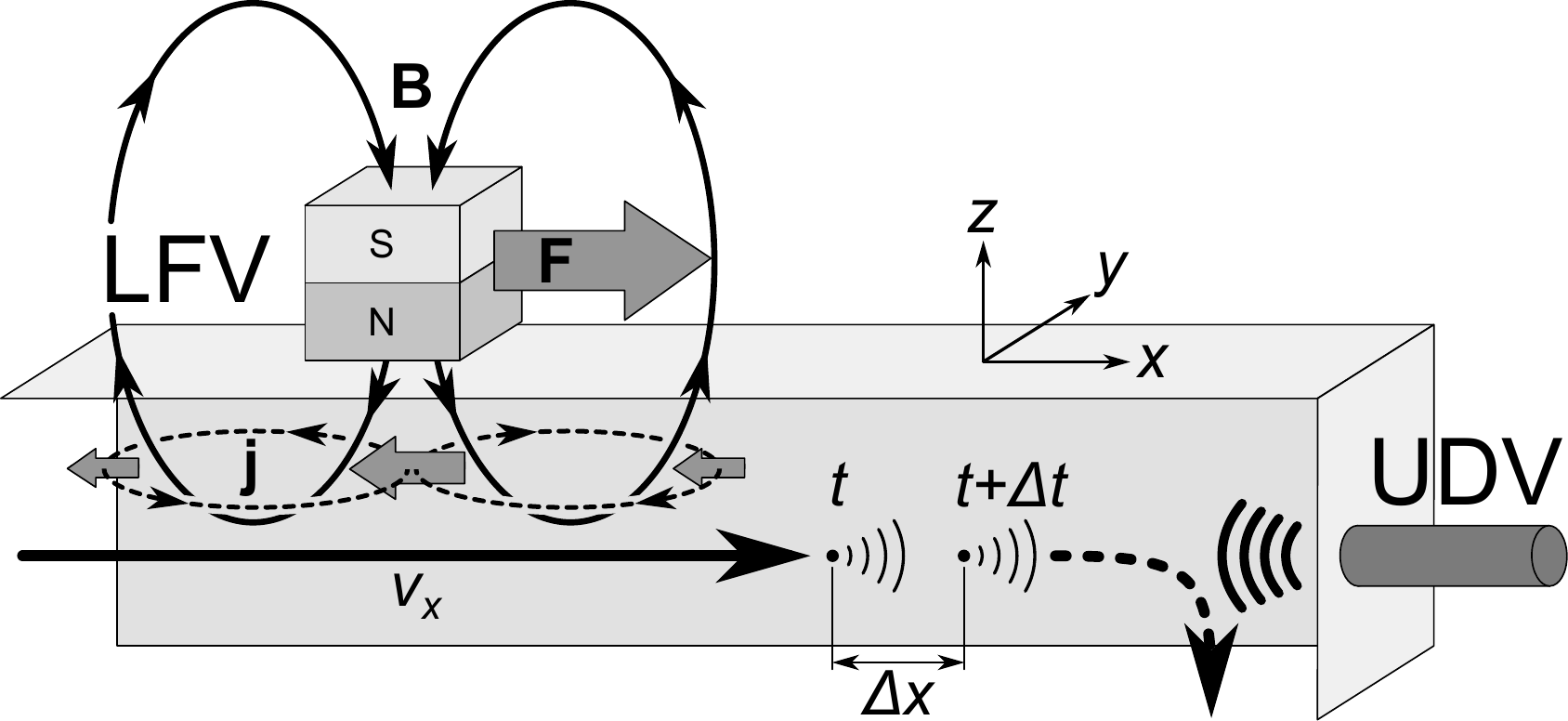}
\caption{
 Schematic of the flow measurement techniques. LFV uses the magnetic field $\vec B$ of a permanent magnet to induce eddy currents $\vec j$ in the flow, which in turn generate forces $\vec F$ in the fluid and on the magnet. UDV sends an ultrasonic burst into the fluid and measures the position of particles at a time $t$ by recording their echoes. After a short time interval $\Delta t$ a second measurement detects a shift in position by $\Delta x$ along the axis of the UDV probe (in this case the $x$-axis). The longitudinal velocity component is then $v_x=\Delta x/\Delta t$.}
\label{fig:meastech}
\end{figure*}
In this work we investigate the applicability of Lo\-rentz force velocimetry (LFV) \citep{Thess2006,Thess2007} to liquid metal convection. Here, the flow is subjected to the outer magnetic field of a permanent magnet, which generates mo\-tion-induced eddy currents in the liquid metal (see Figure~\ref{fig:meastech}). These currents give rise to Lorentz forces in the fluid by interacting with the applied outer magnetic field. The forces are directed opposite to the flow and act as a brake on the fluid motion. At the same time, due to Newton's Third Law, a force in the range of micro- to millinewton acts on the permanent magnet which can be measured by precision methods \citep{Heinicke2012}. It has the same magnitude as the sum of all Lorentz forces in the liquid, but is directed in the opposite direction -- the magnet is in effect dragged along with the flow. The LFV technique does not require any contact with the liquid, which makes it especially interesting for chemically aggressive or hot liquids such as steel melts. The braking effect of LFV on the flow can be neglected for high velocities; the case of slow flows will be addressed later in this work.\par
The Lorentz force $\vec F_L$ in a fluid volume $V$ is given by
\begin{equation}
\label{eq:FL}
\vec F_L = \int_V \vec j(\vec r)\times\vec B(\vec r) \de V \,,
\end{equation}
with $\vec B$ being the magnetic induction  (or magnetic field) and $\vec j$ the current density. The current density itself is connected by Ohm's law to the velocity field $\vec v$ and the magnetic field $\vec B$, namely by $\vec j=\sigma(-\nabla \varphi +\vec v\times \vec B)$ with $\varphi$ being a scalar potential and $\sigma$ being the electrical conductivity of the liquid metal. Dimensional analysis results in the following relation
\begin{equation}
\label{eq:FL_scaling}
F_L \sim \sigma U B^2 V \,,
\end{equation}
where $U$ is a typical flow velocity amplitude, e.g. a root mean square or a mean velocity, and $B=|\vec B |$.\par
This scaling (\ref{eq:FL_scaling}) is valid for the quasistatic approximation of magnetohydrodynamics \citep{Davidson2001}, where the retroactive effect of the induced magnetic field on the eddy currents can be neglected. The linear dependence of the force on the velocity field $\vec v$ has been successfully used, among others, in liquid metal duct flows \citep{Wang2011a}, for electrolytes with weak electrical conductivity \citep{Wegfrass2012,Vasilyan2014, Wiederhold2016a} and for the flow in a rotating tank with significant velocity changes \citep{Sokolov2016}. In the latter two examples the LFV method has been pushed to the limits of applicability, i.e. to a regime where the assumption of the quasistatic approximation breaks down or the liquid exhibits an electrical conductivity that is very small. To maximize the measured force signal most of these experiments have used a magnetic field that penetrates the whole cross-section of the duct and measured the total volume flux. Another approach is to restrict the fluid volume subjected to the magnetic field to a small area. The resulting force on the magnet is then only influenced by the local flow in that volume. This approach is called {\em Local Lorentz Force Velocimetry} (LLFV)  and can be used to probe for example the profile of a liquid metal flow in a duct \citep{Heinicke2013a} or in a continuous casting mould experiment \citep{Hernandez2016a}. The resolution of LLFV is clearly determined by the size of the magnet that probes the induced Lorentz forces. All examples that were mentioned so far have one thing in common. There is a well-defined (mean) flow direction and/or the velocity magnitude is sufficiently large since the momentum transfer into the flow proceeds directly via sustained shear or pressure gradients.\par
The motivation for the present work is twofold. Firstly, we want to explore the applicability of LLFV to thermal (or natural) convection. These flows exhibit in general much smaller Reynolds numbers since they are driven by temperature differences that generate high shear rates via thermal plumes. In our case at hand velocity magnitudes will thus rather be of the order of mm/s than cm/s or m/s. This results via (\ref{eq:FL_scaling}) in much smaller force signals, which make the measurement process as a whole much more challenging. We will also investigate whether the induced Lorentz forces influence the local velocity. In this respect, we want to explore a further limit of this contactless method of velocity measurement in opaque fluids.\par
Secondly, we take this opportunity and measure the turbulent transport laws of heat and momentum in a further liquid metal flow that has not been explored experimentally in this parameter regime. Vertical convection with opposite side walls that are held at a temperature difference $\Delta T$ has recently received a new interest as a further testing case for scaling theories of turbulent transport \citep{Ng2013,Ng2015,Shishkina2016a}. Liquid metals are very good heat conductors which positions them into the class of low-Prandtl-number convection flows. For both reasons, the vertical convection is well suited as a benchmark experiment.\par
Furthermore, we will show that the large-scale flow structure (also known as large-scale circulation or LSC) in this setting remains relatively simple with one mean flow roll that extends across the whole convection cell. This is in stark contrast to the well-known case of Rayleigh-Bénard convection (RBC), where a fluid layer is heated from below and cooled from above. RBC exhibits mostly transient flow structures, which are subject to reversals and cessations \citep{Brown2006,Zhou2009}. This unpredictability makes RBC less practicable for benchmarking our measurement method.\par
Our LLFV measurements are complemented by applying ultrasonic Doppler velocimetry (UDV). In this method an ultrasonic burst is sent into the liquid. The burst is generated by a piezo-crystal in a transducer, which is either in direct contact with the liquid or sends the signal through the wall of the fluid container. The burst travels along the continued centreline of the transducer and is reflected by small particles suspended in the liquid. The returning echo is recorded by the transducer (see Figure~\ref{fig:meastech}). The elapsed time between the emission of the burst and the return of the echo can be converted into a position along the ultrasonic beam by knowing the speed of sound of the liquid. Originally, UDV determined the flow velocity from the Doppler shift of the echo from the original frequency \citep{Takeda1986}. For reasons of fast data processing, this has been changed into a procedure, where multiple successive measurements are correlated and the shift in particle position is converted into the flow velocity. The result is a one-dimensional, one-component velocity profile along the beam axis of the velocity component parallel to the beam. UDV has been successfully applied in a variety of rotating and non-rotating liquid metal flows \citep{Brito2001, Eckert2002, Vogt2013, Vogt2014a, Tasaka2016}.\par
The outline of the article is as follows. Section \ref{sec:setup} will discuss the experimental setup and lists all important parameter definitions. It is followed by a short discussion of typical velocity profiles and time series as well as the LSC flow. Section \ref{sec:lfvudv} summarizes our findings for the LLFV before switching to the global transport laws of heat and momentum in section \ref{sec:scaling}. Finally, we give a brief outlook.
\begin{figure*}[t]
%
\subfigure[\label{fig:VCsetup_Geom}]
  {\includegraphics[height=.3\textwidth]{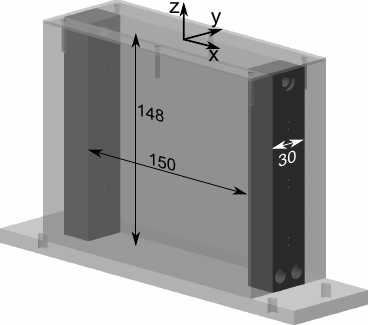}}
\hspace{.1\textwidth}
\subfigure[\label{fig:VCsetup_MeasTech}]
  {\includegraphics[height=.3\textwidth]{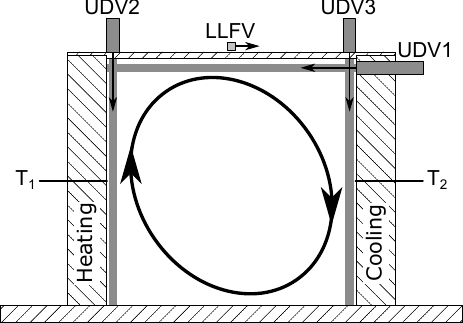}}
\caption{
  Experimental setup. 
  \protect\subref{fig:VCsetup_Geom} Sketch with inner cell dimensions in mm.
  \protect\subref{fig:VCsetup_MeasTech} Side view with sensors and sketch of the LSC for $T_1 > T_2$. The ultrasonic propagation directions are indicated as grey lines.}
\label{fig:setup}
\end{figure*}
\section{Experimental Setup}
\label{sec:setup}
The experiments are conducted in a closed rectangular cell of width 150\,mm, thickness 30\,mm and height 148\,mm (see Figure~\ref{fig:VCsetup_Geom}). The narrow side walls consist of heat exchangers made from copper. They are heated or cooled, respectively, using water from separate thermostats. All other walls are made of PMMA. The cell is filled with the eutectic alloy gallium-indium-tin (GaInSn). Table \ref{tab:GaInSn} summarizes some important material parameters according to \citet{Plevachuk2014}. The coordinate system is positioned at the centre of the top surface with $z$ in vertical upward direction and $x$ pointing horizontally towards the cooling plate.\par
The temperature difference $\Delta T$ between the copper plates is measured using two K-thermocouples at the centre of each copper plate, their tip being in contact with the liquid. $T_1$ is the temperature at the hot plate and $T_2$ at the cold plate. The entire cell is thermally insulated using Styrofoam plates and insulation wool. Additionally a Styrofoam box is placed around the whole experiment to prevent air circulations to influence the force measurement. To determine the heat flux across the cell, two additional K-thermocouples measure the temperature of the in- and outgoing water $T_\text{in}$ and $T_\text{out}$ of the cooling heat exchanger. The volume flux $\dot V$ of the cooling water is measured using an axial turbine flow sensor.\par
The LLFV measurement system \citep{Heinicke2012} consists of a cubic permanent magnet of side length 5\,mm, which is placed on a parallel spring. The deflection of the spring through the force acting on the magnet is measured by a laser interferometer.  The system is placed on top of the cell with the magnet 5\,mm above the liquid and centred at $x = y = 0$\,mm (see Figure~\ref{fig:VCsetup_MeasTech}). The force $F_x$ on the magnet is measured along the $x$-axis, which coincides with the expected flow direction of the LSC at this point. The sampling frequency is 6.3\,Hz, which is the maximal frequency that still results in a linear dynamic response of the parallel spring.\par
The UDV measurements are performed along three lines. We use 8\,MHz transducers with a piezo-element of 5\,mm diameter. The first sensor UDV1 measures the velocity $v_x$ along the $x$-axis, 5.5\,mm below the top surface of the liquid. It is placed in a hole through the cooling cooper plate and is in direct contact with the liquid metal. The second and third sensors, UDV2 and UDV3, are placed on top of the cell, such that each beam line is 4\,mm away from one side wall. They measure $v_z$ along the $z$-axis. Both sensors are installed on the outside of the cell so that the acoustic coupling to the fluid is realized through the 4mm thick top wall. All three sensors are centred in the $y=0$ plane. Simultaneous measurements of multiple sensors are done using a \emph{DOP3010} velocimeter and measurements of single sensors utilize a \emph{DOP2000} velocimeter by \emph{Signal Processing SA}. The spatial resolution along the propagation direction is $\lesssim0.35$\,mm. The time resolution depends on the number of pulses that are used to calculate one velocity profile and the frequency of the pulse emission. The latter is called the pulse repetition frequency (PRF) and is set to 500\,Hz. For joint measurements of LLFV and UDV the time resolution is 0.64\,s with 300 emissions per profile. If UDV is used alone, the time resolution is 0.54\,s with 250 emissions per profile.\par
\begin{table}[t]
\caption{Properties of eutectic GaInSn at 25 $^\circ$C \citep{Plevachuk2014}.}
\label{tab:GaInSn}
\begin{tabular}{llcl}
\hline\noalign{\smallskip}
Composition percentage Ga & & & 67.0 wt-\% \\
Composition percentage In & & & 20.5 wt-\% \\
Composition percentage Sn & & & 12.5 wt-\% \\
Mass density & $\rho$ &$=$& $6.3\times 10^3$\,kg/m$^3$ \\
Kinematic viscosity & $\nu$ &$=$& $3.3\times 10^{-7}$\,m$^2$/s \\
Thermal diffusivity & $\kappa$ &$=$& $1.0\times 10^{-5}$\,m$^2$/s \\
Isobaric heat capacity & $c_p$ &$=$& 365\,J/(kg K) \\
Electrical conductivity & $\sigma$ &$=$& $3.2\times 10^6$\,S/m \\
Volumetric expansion coefficient & $\alpha$ &$=$& $1.2\times 10^{-4}$\,1/K \\
\noalign{\smallskip}\hline
\end{tabular}
\end{table}
From these measurements the following dimensionless numbers are derived, using the thermophysical properties of GaInSn \citep{Plevachuk2014} at the mean temperature $T_0 = (T_1+T_2)/2$. The Rayleigh number $Ra$ is calculated from the measured temperature difference $\Delta T = T_1-T_2$ and the cell width $L=150$\,mm. It is given by 
\begin{equation}
Ra = \frac{\alpha g\Delta T L^3}{\nu\kappa} \,,
\end{equation}
with $\alpha$, $\nu$ and $\kappa$ being the volumetric expansion coefficient, the kinematic viscosity and the thermal diffusivity of GaInSn, respectively. The variable $g$ stands for the acceleration due to gravity. The second important parameter is the Prandtl number $Pr$, which is given by 
\begin{equation}
Pr=\frac{\nu}{\kappa}\approx 0.033\,.
\end{equation}
The Nusselt number $Nu$ is the quotient of the total heat flux~$\dot Q$ through the cell, compared to the purely diffusive heat flux~$\dot Q_\kappa$. Neglecting any heat losses to the surrounding, the total heat flux is equal to the heat received by the cooling water of the heat exchanger:
\begin{align}
\dot Q &= \tilde c_p\tilde\rho\dot V (T_\text{out}-T_\text{in}) \,.
\end{align}
$\tilde c_p$ and $\tilde\rho$ are the specific heat and mass density of water \citep{Cengel2008}. The diffusive heat flux is given by 
\begin{align}
\dot Q_\kappa &= \kappa c_p\rho A \frac{\Delta T}{L} \,,
\end{align} 
where $c_p$ and $\rho$ are the specific heat at constant pressure and mass density of GaInSn and $A = (148 \times 30)\,\text{mm}^2$ is the cross section of the cell. Thus we get
\begin{align}
Nu &= \frac{\dot Q}{\dot Q_\kappa} 
  = \frac{\tilde c_p\tilde\rho}{\kappa c_p\rho} \frac{\dot VL}{A} 
    \frac{T_\text{out}-T_\text{in}}{\Delta T} \,.
\end{align}
The Reynolds number $Re$ is calculated from the one-di\-men\-sio\-nal velocity profiles of the UDV-measurement. For every time~$t$ a characteristic horizontal velocity~$U_x$ is derived: The absolute velocities measured by UDV1 are averaged over the interval $x\in[-40,+40]$\,mm. Similarly, a vertical characteristic velocity~$U_z$ is calculated from the velocities recorded by UDV2 and UDV3 in the interval $z \in [-115, -35]$\,mm. Thus, 
\begin{align}
\label{eq:charVelo_horizontal}
U_x(t) &= \bigl\langle|v_x(x,t)|\bigr\rangle_{x\in[-40,+40]\text{\,mm}} \,,\\
\label{eq:charVelo_vertical}
U_z(t) &= \bigl\langle|v_z(z,t)|\bigr\rangle_{z\in[-115, -35]\text{\,mm}} \,.
\end{align}
In these intervals, we expect the direction of the LSC to be generally parallel to the measurement axis of the respective sensor (see also Figure~\ref{fig:LSC_UDV}). A global characteristic velocity $U$ is calculated by using the velocities of all three sensors in their respective intervals. However, the horizontal sensor UDV1 is counted twice in this average, since we have two vertical sensors but only one horizontal sensor (this emulates an additional horizontal sensor along the bottom of the cell). These three characteristic velocities are then used to calculate a horizontal, vertical and global Reynolds number 
\begin{align}
\label{eq:ReynoldsNumbers}
Re_x &= \frac{U_x L}{\nu} \,, &
Re_z &= \frac{U_z L}{\nu} \,, &
Re &= \frac{U L}{\nu} \,,
\end{align}
respectively.
\section{Large-scale circulation and velocity statistics}
\label{sec:lsc}
\begin{figure}[t]
\includegraphics[width=\columnwidth]{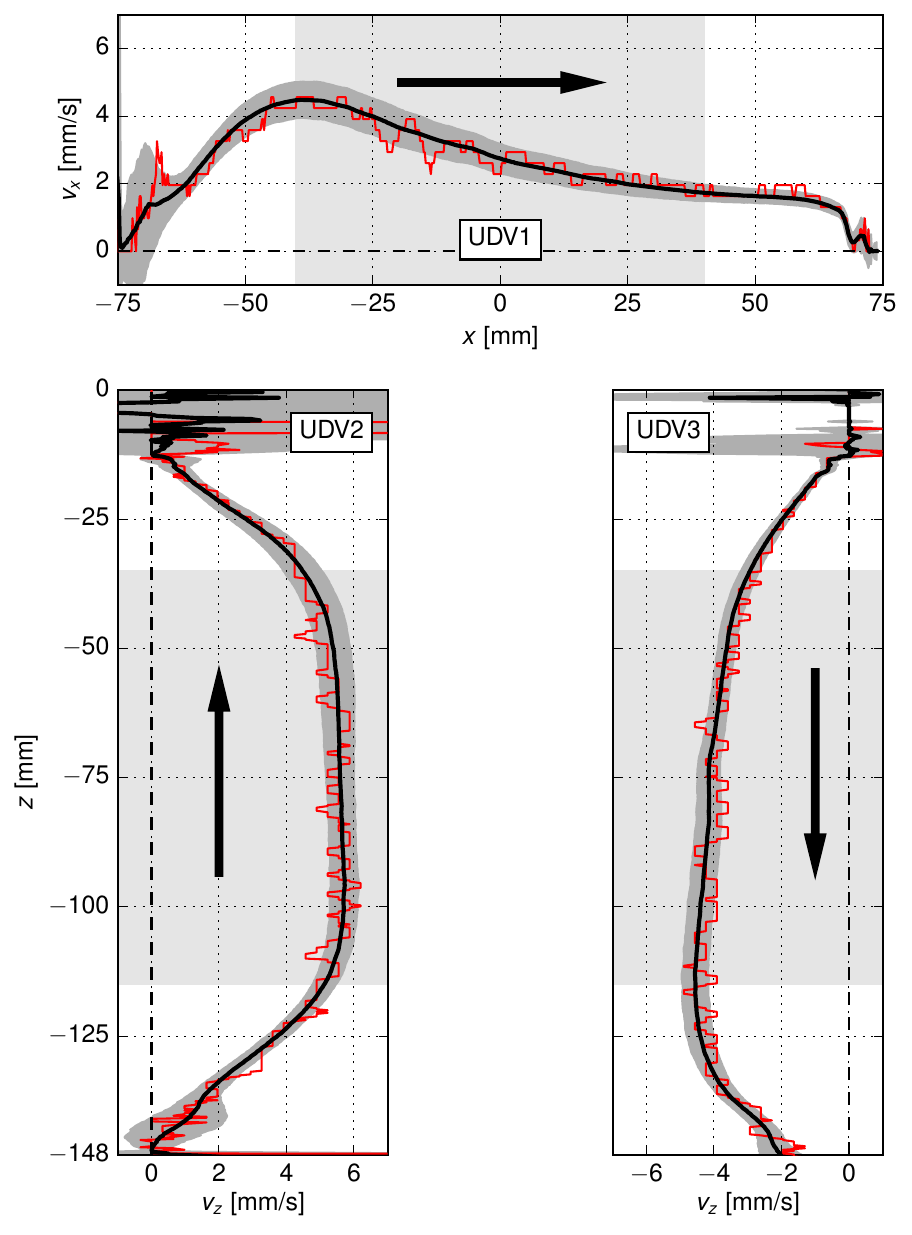}
\caption{
  Velocity profiles measured by the UDV probes at $Ra = 1.6\times10^6$. Thick black lines: Mean velocity profile over 1770 snapshots. Thin red lines: Typical velocity snapshot. Dark grey envelope: Standard deviation from the mean profile. Light grey areas: Depth-intervals for the calculation of characteristic velocities and probability density functions.}
\label{fig:LSC_UDV}
\end{figure}
\begin{figure}[t]
\includegraphics[width=\columnwidth]{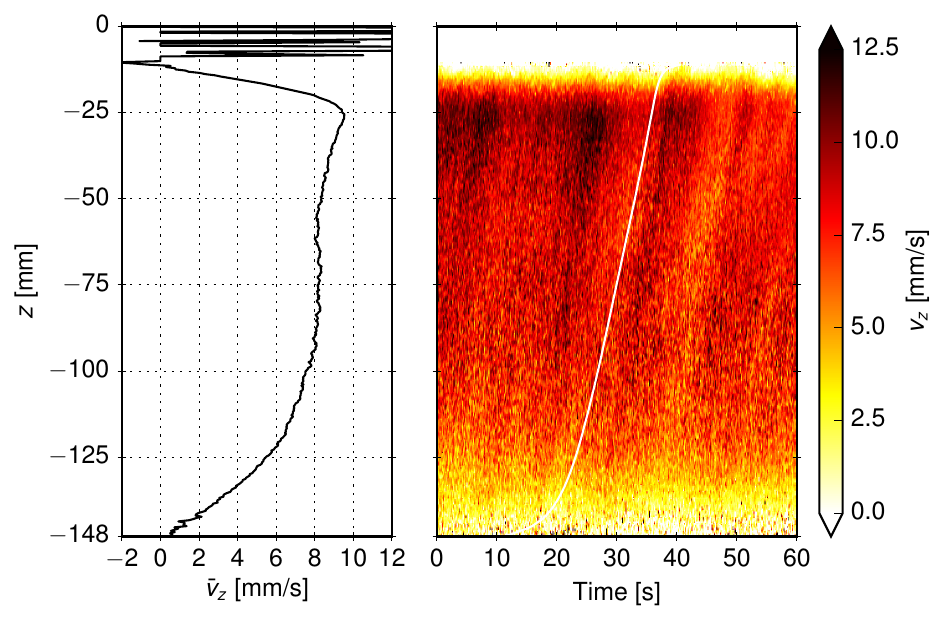}
\caption{
  Time averaged velocity profile $\bar v_z(z)$ (left) of the velocity field $v_z(z,t)$ (right) measured by UDV2 at $Ra=6.2\times10^6$. The right contour plot gives the colour-coded velocity over time and position. The white line in the right plot is the trajectory~\eqref{eq:particleTrajectory} of a particle moving with $\bar v_z(z)$ across the cell. The top 10~mm of the velocity field are omitted from the contour plot.}
\label{fig:LSC_patternMovement}
\end{figure}
In Rayleigh-Bénard convection (RBC) a flow only arises when the Rayleigh number exceeds a critical value. Below that point any perturbation of density stratification is stabilised by dissipative forces due to kinematic viscosity and thermal conduction in the fluid. This is not the case for vertical convection. Even the smallest temperature difference between opposing side walls triggers a convective flow \citep{Batchelor1954}. The hot fluid near the heated plate has a lower density than the cold fluid on the other side of the cell. This density gradient generates buoyancy forces and the fluid on the hot side rises up, while it sinks down on the opposite side of the cell. These up- and downwelling flows hit the top or bottom of the cell, respectively, and are redirected in the horizontal direction. They finally combine into one coherent circulation, the LSC, across the whole cell, which is the dominant flow feature of vertical convection, in particular in a closed cell of aspect ratio~1. The flow direction is canonically prescribed without cessations and reversals as known from RBC \citep{Brown2006,Zhou2009}.\par
Figure~\ref{fig:LSC_UDV} shows the time averaged velocity profiles measured by the UDV probes for $Ra = 1.6\times 10^6$. The directions of the flow (indicated by arrows) confirm the existence of the LSC: We see a positive $v_z$ component near the hot wall (UDV2) and negative $v_z$ values for the cooling plate (UDV3). The horizontal flow near the top (UDV1) flows from the hot to the cool side of the cell and closes the circulation.\par
It has to be mentioned that sensors UDV2 and UDV3, which measure indirectly through the cell lid, have a significant dead zone close to the sensor, where the signal is unusable due to excessive noise. This is caused by the formation of multiple acoustic echoes within the lid. These strong echoes have to decay first, before the much weaker signals from the particles in the fluid can be detected. In our case this makes the UDV2 and UDV3 signals unusable for $z \gtrsim -15$\,mm. The UDV1 sensor is in direct contact with the liquid metal and has a much smaller dead zone ($\lesssim 5$\,mm). This is unavoidable due to the ringing of the piezo crystal in the sensor.\par
Each of the three averaged velocity profiles in Figure~\ref{fig:LSC_UDV} are plotted together with an exemplary profile from a single snapshot. In addition to the random fluctuations present in these snapshots, there are persistent flow structures of higher or lower speed than the mean flow. They can be seen as slanted lines in Figure~\ref{fig:LSC_patternMovement}. These structures move roughly with the mean velocity of the flow: A fluid element that moves with the time-averaged velocity $\bar v_z(z) = \langle v_z(z,t)\rangle_t$ across the cell has the trajectory $(z,t(z))$ with
\begin{align}
\label{eq:particleTrajectory}
t(z) &= t_0 + \int_{z_0}^{z}\frac{\de z'}{\bar v_z(z')} \,.
\end{align}
Here $z_0$ and $t_0=t(z_0)$ are the starting position and time, respectively. The trajectory is plotted in Figure~\ref{fig:LSC_patternMovement} as a white line. It matches closely the angle of the patterns in the velocity field. That means, these flow structures are transported by the mean velocity of the flow.\par
\begin{figure}[t]
\includegraphics[width=\columnwidth]{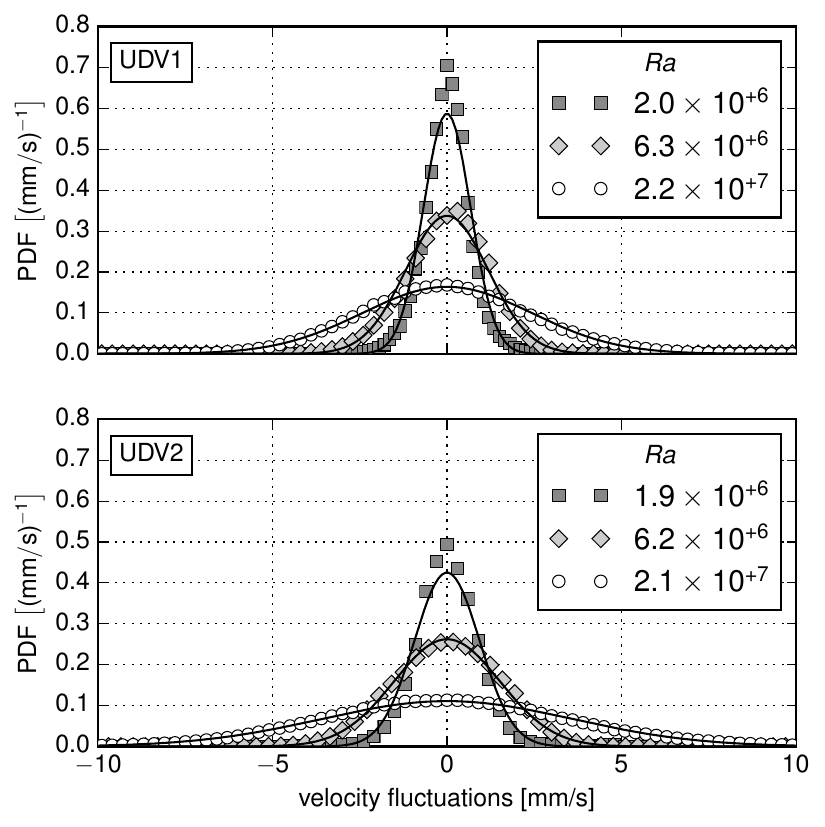}
\caption{
  Probability density function (PDF) of velocity fluctuations measured by UDV1 (top) and UDV2 (bottom). The black lines are normal distributions with the standard deviation of each PDF (see table~\ref{tab:LSC_PDFstats}) and a mean of zero.}
\label{fig:LSC_PDF}
\end{figure}
\begin{table}
\caption{
  Statistical values of the PDFs in Figure~\ref{fig:LSC_PDF}. The relative standard deviation (RSTD) is normalized by the mean value. The excess kurtosis in the last column is defined as the standardized fourth-order moment minus a value of 3 for the normal or Gaussian case.}
\label{tab:LSC_PDFstats}
\begin{tabular}{rrrrrr}
\hline\noalign{\smallskip}
\multicolumn{1}{l}{} & 
\multicolumn{1}{l}{$Ra$} & 
\multicolumn{1}{l}{Mean} & 
\multicolumn{1}{l}{RSTD} & 
\multicolumn{1}{l}{Skewness} & 
\multicolumn{1}{l}{Excess} \\ 
&$\left[10^6\right]$& [mm/s] &&& \multicolumn{1}{l}{kurtosis} \\
\noalign{\smallskip}\hline\noalign{\smallskip}
UDV1 & $2.0$ & $2.67$ & $0.25$ & $-0.094$ & $1.713$\\
     & $6.3$ & $7.75$ & $0.15$ & $-0.176$ & $0.491$\\
     & $21.7$ & $17.20$ & $0.14$ & $0.122$ & $0.050$\\
UDV2 & $1.9$ & $4.61$ & $0.20$ & $-0.040$ & $5.068$\\
     & $6.2$ & $8.67$ & $0.18$ & $-0.078$ & $-0.217$\\
     & $21.5$ & $19.63$ & $0.18$ & $0.043$ & $0.141$\\
\noalign{\smallskip}\hline
\end{tabular}
\end{table} 
In the following, we want to investigate the statistical properties of the fluctuations around the mean velocity profile. Figure~\ref{fig:LSC_PDF} shows the probability density functions (PDF) of the velocity fluctuations. The PDFs are calculated from $1.6\times10^7$ samples measured over 2.6 hours by the UDV1 and UDV2 sensors (here the time resolution of the UDV measurement was decreased to 0.14\,s with 50 emissions per profile). Again, only velocities from the central depth intervals were used (see Figure~\ref{fig:LSC_UDV}). The fluctuations are calculated around the time average for every position separately. Table~\ref{tab:LSC_PDFstats} lists the statistical properties of the PDFs. Additionally, normal distributions with the standard deviation of each PDF are plotted in Figure~\ref{fig:LSC_PDF}.\par
For increasing $Ra$ the PDFs get closer to the shape of a normal or Gaussian  distribution. In particular the excess kurtosis approaches zero (and thus the value of a normal distribution) from initially large values: At low $Ra$ the fluctuations drop off faster, than for high $Ra$. While the standard deviation (STD) increases, the relative STD (RSTD), normed by the velocity mean, vary only slightly except for the lowest $Ra$. The skewness does not show any particular trends. The changes in its values are more likely a sign of a still insufficient sample size in order to determine this specific odd-order moment.\par
In conclusion, we can confirm by UDV measurements that the basic flow structure is one convection roll spanning the whole cell and persisting for all $Ra$. The velocity fluctuations grow linearly with the average speed, but approach a normal or Gaussian distribution for increasing $Ra$. This shows, that in the investigated $Ra$-range we transition from a non-linear flow regime to fully developed turbulence.
\section{Local Lorentz force velocimetry}
\label{sec:lfvudv}
\begin{figure}[t]
\includegraphics[width=\columnwidth]{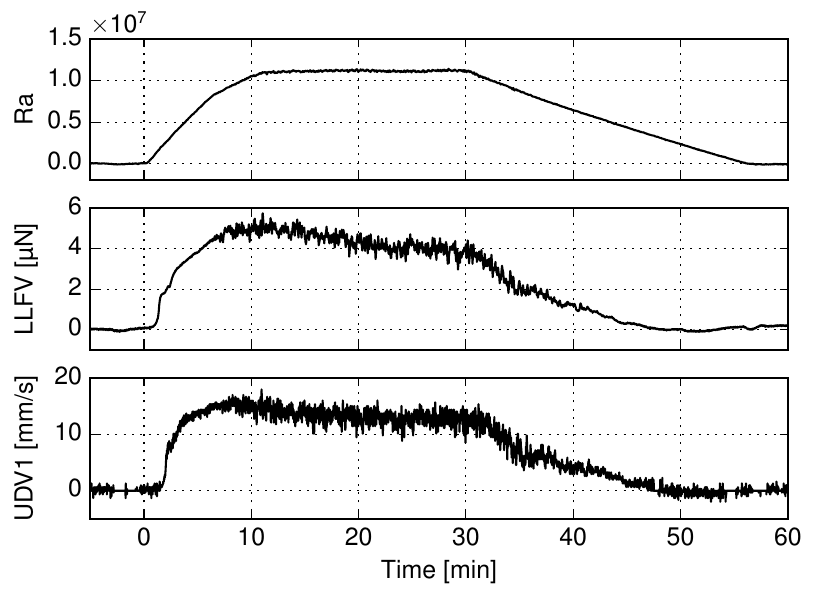}
\caption{
  Time series of experimental data for $Ra=1.1\times10^7$. Top: Rayleigh number. Middle: Force signal $F_x$ of the LLFV system. Bottom: $v_x$ at $x=0$\,mm, measured by UDV1. At $t=0$\,min the the bottom plate is heated up to $Ra = 1.1\times10^7$ and cooled back down to $Ra=0$ at $t=30$\,min. The force and velocity signals settle after $t\sim20$\,min.}
\label{fig:lfvudv_timeSeries}
\end{figure}
For the comparison of UDV and LLFV measurements, experiments were performed at different $Ra$. Figure~\ref{fig:lfvudv_timeSeries} shows an exemplary time series for an experimental run at $Ra=1.1\times10^7$. First, both sides of the cell were set at the same temperature and the zero signal of the force sensor was measured. Then, the temperature of the heating thermostate was raised to set the desired Rayleigh number. Once a stable temperature distribution in the cell was reached, the LLFV and UDV signals were recorded for about ten minutes. Subsequently, the heating temperature was lowered to the initial state and a second zero signal was recorded. The two zero measurements allowed us to correct any linear drifts in the force signal. In these experiments we investigated Rayleigh numbers in the range of $Ra$ from $4\times10^5$ to $3\times10^7$.\par
\begin{figure}[t]
\includegraphics[width=\columnwidth]{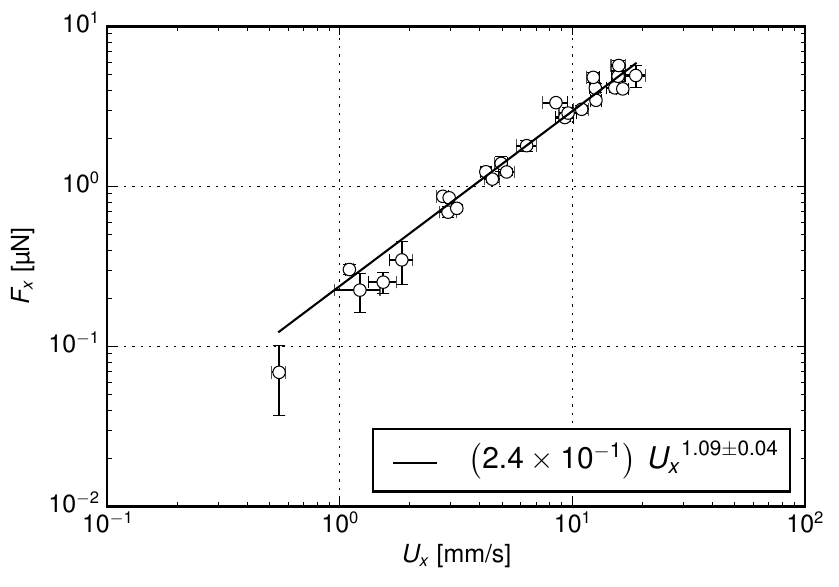}
\caption{Comparison of the characteristic horizontal velocity~$U_x$ (see eq.~\eqref{eq:charVelo_horizontal}) and force~$F_x$ (LLFV) for $Ra= 4\times10^5$ to $3\times10^7$.}
\label{fig:lfvudv_Fx(Ux)}
\end{figure}

Figure~\ref{fig:lfvudv_Fx(Ux)} shows the dependence of the horizontal force $F_x$ of the LLFV on the characteristic horizontal velocity~$U_x$ (see eq.~\eqref{eq:charVelo_horizontal}) measured by the UDV1 sensor just below the top of the cell. For velocities of the order of $10$\,mm/s we measured forces of $\sim 4$\,\textmu N. A power-law fit to the data using orthogonal direction regression shows, that the force $F_x \propto U_x^{1.09}$ is close to a linear scaling with $U_x$. If data for $U_x > 7$\,mm/s are used only, the exponent changes to $1.03 \pm 0.25$. This result is consistent with the expectations from all previous studies of LLFV. It shows that LLFV is sensitive enough even for such low velocities and thus proves the applicability of LLFV in convection flows.\par
Difficulties arise however when LLFV is used for long-term measurements of several hours. For such low-mag\-ni\-tude forces, we saw drifts in the signal which were of the same magnitude as the measured forces. For short periods  these drifts were generally linear and could thus be compensated by zero measurements as described above. However during longer experiments, running for several hours, these drifts can vary in time, which prohibits a proper compensation using zero measurements before and after the experiment. Multiple reasons for these drifts can be given. For example, parasitic electromagnetic fields from surrounding devices, very small shifts in alignment to the vertical axis and, particularly in an experiment driven by temperature differences, the change of the surrounding air temperature can have an influence on the characteristics of the force sensor. With so many environmental influences it was not possible to consistently identify and disentangle any single cause for these signal drifts. While it is possible to use LLFV effectively with forces in the range of \textmu N, as has been done by e.g. \citet{Wiederhold2016a}, it is for now limited to shorter-term measurements.\par
A sufficiently strong amplitude of the Lorentz force will influence the local fluid motion. This effect is well known and is utilized in flow control of liquid metals \citep{Davidson2001,Asai2012}. To quantify the influence of the magnetic field on the flow, we calculate the interaction parameter which is given by
\begin{align}
N &= \frac{\sigma B^2 l}{\rho U_x} \,.
\end{align}
Quantity $B$ is the maximal field strength in the fluid, in our case 5\,mm away from the magnet surface. This value was measured using a Gaussmeter to be $B=63$\,mT. The scale~$l$ is a characteristic length of the magnetic field in the liquid. For this we estimated the penetration depth of LLFV to be $l=5.7$\,mm (see appendix~\ref{sec:penetration} for further details). For $N \ll 1$ the deformation of the flow field by the induced Lorentz forces can be neglected. However, once $N$ reaches or exceeds unity, the flow may be altered. Since natural convection exhibits low velocities and $N\propto 1/U_x$ this potential impact on the flow has to be investigated.\par
\begin{figure}
\includegraphics[width=\columnwidth]{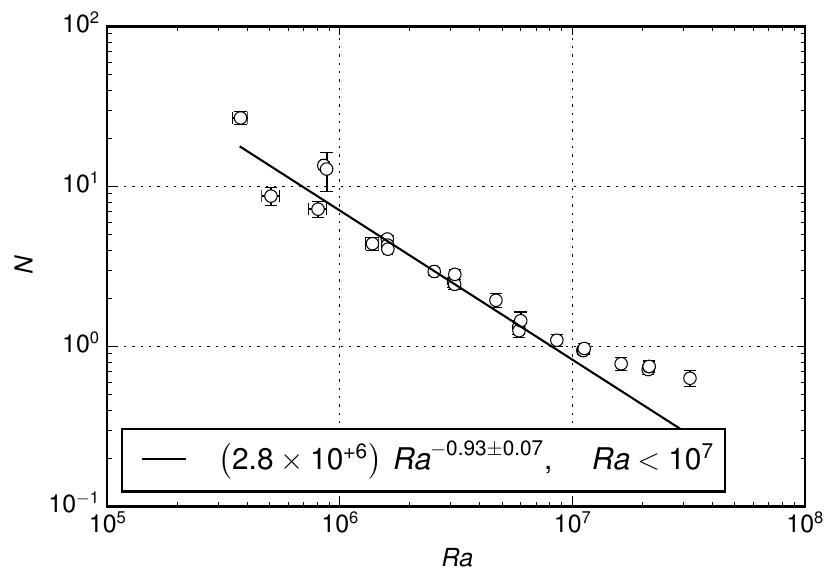}
\caption{
  Dependence of the interaction parameter $N$ on the Rayleigh number $Ra$. A power-law, which is indicated by the solid line, was fitted to the points with $Ra < 10^7$.}
\label{fig:lfvudv_N(Ra)}
\end{figure}
\begin{figure}
\includegraphics[width=\columnwidth]{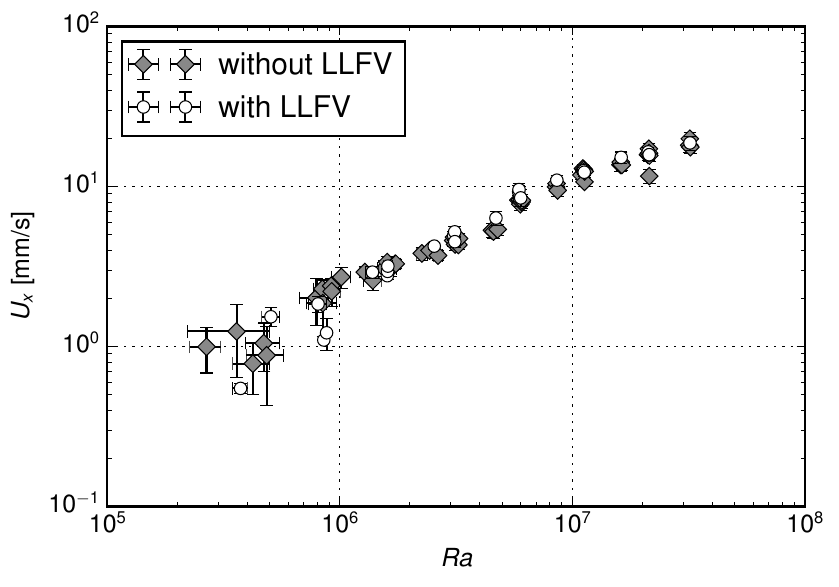}
\caption{
  Comparison of the characteristic horizontal velocities~$U_x$ (see eq.~\eqref{eq:charVelo_horizontal}) measured by UDV1 with and without the presence of the magnetic field of the LLFV system.}
\label{fig:lfvudv_Ux(Ra)}
\end{figure}

Figure \ref{fig:lfvudv_N(Ra)} shows, that $N > 1$ for $Ra < 10^7$. At the threshold of $N=1$ we can also see a change in the scaling of $N(Ra)$. This scaling can be linked directly to the flow velocity, since $N\propto 1/U_x$. However, when comparing $U_x$ for the cases with and without the influence of the magnetic field from the LLFV system in Figure~\ref{fig:lfvudv_Ux(Ra)}, there is no significant deviation in this range of $Ra$.\par
The reason that there is no visible change in scaling for the $F_x(U_x)$ relation in Figure~\ref{fig:lfvudv_Fx(Ux)} is, that $U_x$ is measured in-situ. That means, the LLFV measures the resulting velocity that is actually present, no matter whether it is altered by the probing magnetic field or not. In case of forced convection, where the characteristic velocity is prescribed, one can expect to see a deviation of the force scaling for $N>1$; the flow speed near the LLFV sensor would then be altered and not match the prescribed velocity any more. Clearly, LLFV is limited here, keeping in mind, that this method was originally designed for integral flow measurements.
\section{Scaling laws of turbulent heat and momentum transfer}
\label{sec:scaling}
We now examine the behaviour of the transport of heat and momentum by the convective flow. The results presented in this section were recorded without the presence of the LLFV measurement system and the accompanying magnetic field since long-term experimental runs were required. At the beginning of the experiments the cooling and heating thermostats were set to the same temperature for a zero measurement. Then the heating temperature was raised stepwise to establish multiple temperature differences $\Delta T$ across the fluid. Each experiment at a given $\Delta T$ was conducted for about one hour. The dimensionless numbers $Ra$, $Nu$, $Re$, $Re_x$ and $Re_z$ were determined as described in section~\ref{sec:setup}. With a cooling temperature of 15\,$^\circ$C and a maximum heating temperature of 63\,$^\circ$C we were able to cover a range of $Ra = 3 \times 10^5$ to $3 \times 10^7$, i.e. two orders of magnitude. Errors are given as standard deviations. Power law fits use orthogonal distance regression to account for uncertainties in both quantities on the abscissa and ordinate.
\subsection{Heat transport}
\label{sec:heattransport}
\begin{figure}[t]
\includegraphics[width=\columnwidth]{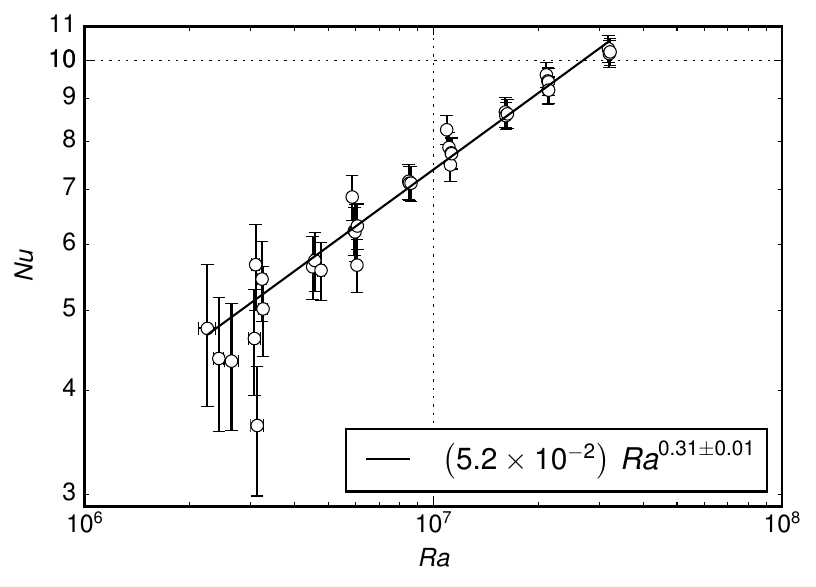}
\caption{Scaling of Nusselt number with Rayleigh number. The solid line indicates the power law fit.}
\label{fig:scal_Nu(Ra)}
\end{figure}
Figure~\ref{fig:scal_Nu(Ra)} shows the dependence of the Nusselt number $Nu$ on the Rayleigh number $Ra$. We only display results for $Ra > 2\times 10^6$, because for lower $Ra$ the temperature difference of the in- and outgoing cooling water was smaller than the accuracy of the temperature measurement. A power-law fit to the data results in a scaling of $Nu \propto Ra^{0.31}$. The same exponent was found by multiple DNS simulations \citep{Ng2015,Yu2007}, even though these were conducted for air ($Pr = 0.71$). The different $Pr$ in the simulations and our experiment lead only to higher absolute values of $Nu$ in the simulations, but the scaling is the same. The exponent of $0.31$ was explained by \citet{Ng2015} as a superposition of 1/4 and 1/3 scaling laws, which can be derived theoretically for the laminar \citep{Shishkina2016a} and turbulent case \citep{Ng2013}, respectively. To conclude this paragraph, our findings are consistent with those from numerical simulations of vertical convection. Interestingly, the scaling exponent of RBC in a liquid metal flow at $Pr=0.021$ is found to be smaller with values of about 0.26, while at $Pr=0.7$ the exponent is 0.29 \citep{Scheel2016}.
\subsection{Momentum transport}
\label{sec:momentumtransport}
\begin{figure}[t]
\includegraphics[width=\columnwidth]{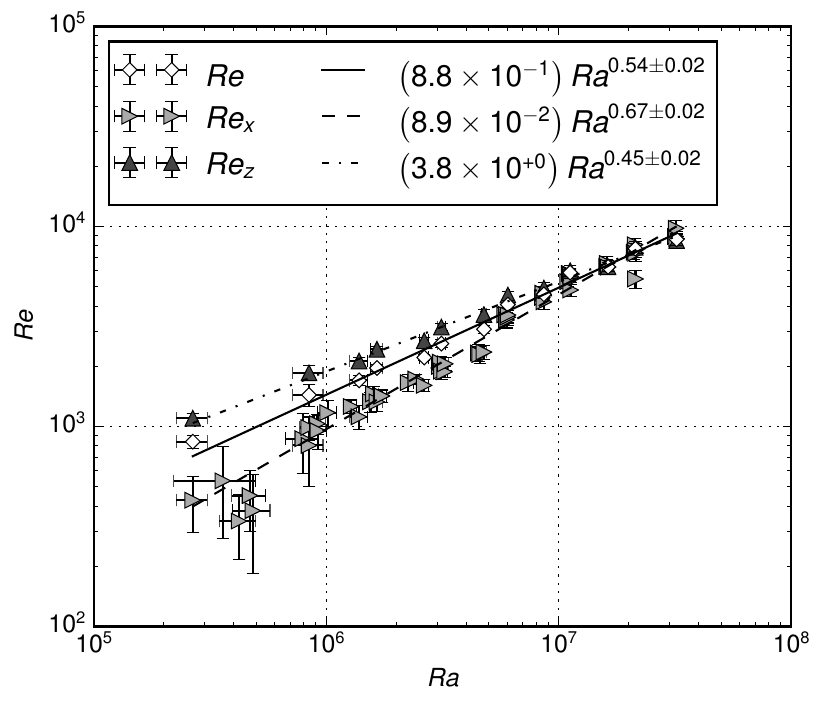}
\caption{
  Scaling of the three Reynolds numbers with Rayleigh number: Global $Re$ (diamonds, solid line), horizontal $Re_x$ (triangles right, dashed line) and vertical $Re_z$ (triangles up, dash-dotted line).}
\label{fig:scal_Re(Ra)}
\end{figure}

The scaling of $Re$, $Re_x$ and $Re_z$ with $Ra$ is displayed in Figure~\ref{fig:scal_Re(Ra)}. The global Reynolds number $Re$ follows a power law of $Re \propto Ra^{0.54}$. This is a combination of the different behaviours of the vertical and horizontal flows in the cell.\par
For the vertical Reynolds number $Re_z$ we see a scaling of $Re_z\propto Ra^{0.45}$. This is close to a 1/2-scaling as found in previous numerical simulations by \citet{Shishkina2016a}, where a maximum vertical velocity was used to calculate a Reynolds number. The deviation in the exponent might stem from the averaging effect over the cross-section of the ultrasonic beam. This result is also very close to RBC in liquid metal flow in the direct numerical simulations by \citet{Scheel2016}.\par
The horizontal Reynolds number $Re_x$ instead follows a $Re_x \propto Ra^{0.67}$ power law, which is very close to an exponent of 2/3. So far, little attention has been given to this velocity component, partly because most numerical studies employ periodic boundary conditions in vertical direction, instead of simulating a closed cell \citep{Ng2013,Ng2015}.\par
The absolute values of $Re_z$ are generally larger than $Re_x$: The fluid is accelerated vertically in a thin layer near the copper plates. Once it reaches the top or bottom of the cell, the flow is redirected in horizontal direction and widens to a broader layer. Since the horizontal motion is driven by the vertical acceleration, this widening reduces the flow velocity. However, because of the stronger growth of $Re_x$ with $Ra$ these velocities converge to a common value. Figure~\ref{fig:scal_Re(Ra)} shows, that $Re_x\sim Re_z$, when $Ra > 10^7$. We expect that $Re_x$ loses its $Ra^{0.67}$ scaling past this point and instead follows the same scaling law as $Re_z$.
\section{Conclusion}
\label{sec:conclusion}
In the present study, we investigated the behaviour of vertical convection in a liquid metal. Local Lorentz force velocimetry and ultrasound Doppler velocimetry were used to measure the flow structure, which consists of a single large scale convection roll. Velocity fluctuations are found to be transported by the large scale circulation (see Figure~\ref{fig:LSC_patternMovement}) and approach a normal (or Gaussian) probability density function for increasing $Ra$.\par
By a direct comparison of LLFV and UDV measurements, the linear response of LLFV to low velocity flows was confirmed. Even though the interaction parameter $N$ in the present work partly exceeded unity, a comparison with the undisturbed flow showed little deviation. However, this may change for higher $N$ or other flows and has to be considered carefully in every application. We nonetheless showed that the liquid metal convection flow is accessible by LLFV and thus a further contactless measurement method is available. Our analysis demonstrated also that particularly shorter-term measurements would be appropriate, which are required in many of the potential applications. One has to keep in mind that small velocity magnitudes are translated into forces of the order of micronewtons in a high-precision force measurement system and that the LLFV system has to be in close proximity to the liquid. Possible extensions of LLFV to so-called time-of-flight measurements by the usage of two identical probes \citep{Dubovikova2016} or arrays of probes would be possible and could reduce the numerous systematic error sources. It is thus clear that we have explored a further limit of LLFV\par
We also studied the turbulent transport properties in vertical convection. The global heat transport follows a scaling law $Nu \propto Ra^{0.31}$ for a range of Rayleigh numbers of $2\times10^6 < Ra < 3\times 10^7$. The momentum transport in vertical direction scales as $Re_z\propto Ra^{0.45}$ and the horizontal momentum transport as $Re_x\propto Ra^{0.67}$ for $3\times 10^5 < Ra < 3\times 10^7$. The resulting global Reynolds number has a dependence of $Re\propto Ra^{0.54}$. These power laws agree well with previous numerical investigations of vertical convection in fluids of higher Prandtl number such as air. Given that the simulations and experiments have been conducted in different geometrical settings, we can conclude that the scaling in vertical convection seems to be less sensitive with respect to geometry effects and Prandtl number. Further vertical convection experiments will however be necessary to substantiate this conclusion. The study of heat and momentum transport of thermal convection in liquid metals as low-Prandtl-number fluids in general promises a better understanding of transport mechanisms and may help to refine theoretical models. \par
The high electrical conductivity also allows in principle the local manipulation of convective flows by external magnetic fields. While this work focussed on LLFV as a measurement method, which ideally leaves the flow unchanged, stronger magnetic fields can alter the flow structure and in turn the transport properties. Particularly for Rayleigh-Bé\-nard convection, with its much more complex flow structures than vertical convection, this can lead to substantial changes in the flow structure. Such insights can be used for flow control of liquid metal flows in the presence of parasitic magnetic fields or by explicitly applying magnetic fields.
\begin{acknowledgements}
TZ is supported by the Research Training Group on Lorentz Force Velocimetry and Lorentz Force Eddy Current Testing which is funded by the Deutsche Forschungsgemeinschaft with grant No. GRK 1567. TV is supported by the LIMTECH Alliance of the Helmholtz Association. We thank Ronald du Puits, Vladimir Galindo, Christian Karcher for discussions and Alexander Thieme for the technical support in the experiments.
\end{acknowledgements}
\appendix
\section{The penetration depth of LLFV}
\label{sec:penetration}
\begin{figure}[h]
\includegraphics[width=\columnwidth]{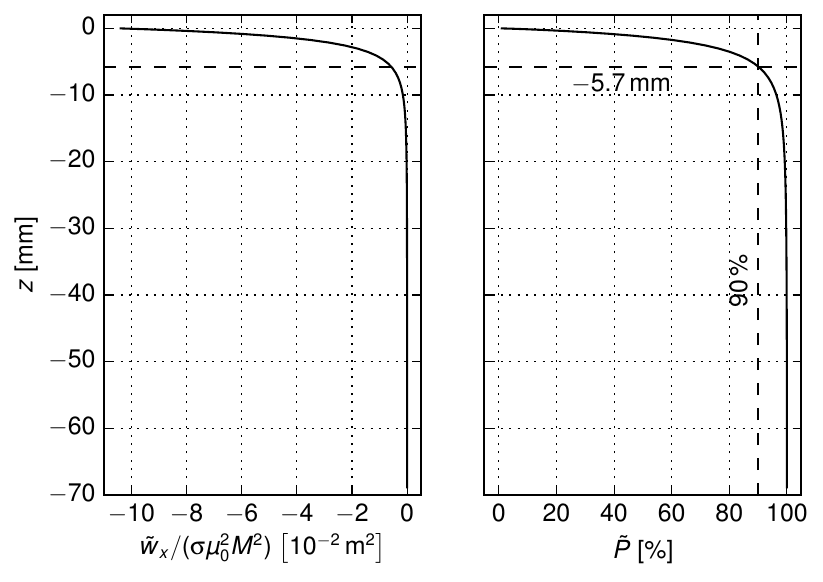}
\caption{
  Weight-function $\tilde w_x(z)$ normalized by $\sigma\mu_0^2 M^2$ (left) and cumulative relative contribution $\tilde P(z)$ (right) for $2l=5$\,mm and $h=7.5$\,mm.}
\label{fig:LFVpen_result}
\end{figure}

An important question is how far LLFV can reach into the liquid, i.e. up to what depth the fluid velocity is influencing the measurement. We calculate this penetration depth for an infinite half-space $V=\{\vec r\in\mathbb R^3: z\le0\}$ filled with liquid metal under the quasistatic approximation. We assume a stationary one-dimensional flow field $\vec v(\vec r) \equiv v_x(z) \vec e_x$, which is probed by the magnetic field $\vec B(\vec r)$ of a permanent magnet outside $V$. The Lorentz force $\vec F_L$ acting on the whole fluid is given by \eqref{eq:FL}. Inserting Ohm's law $\vec j = \sigma(-\nabla\varphi + \vec v\times\vec B)$ gives
\begin{multline}
\label{eq:LFVpen_force1}
\vec F_L = 
  \sigma\int_V \bigl(\vec v(\vec r)\times\vec B(\vec r)\bigr)
    \times\vec B(\vec r) \de V \\
  + \sigma\int_S \vec\varphi(\vec s)\vec B(\vec s)\times\vec n(\vec s)\de S \,.
\end{multline}
Here we used Stokes' theorem and $\nabla\times(\varphi\vec B) = \nabla\varphi \times \vec B$ in $V$. $S=\{\vec s\in\mathbb R^3: z=0\}$ is the surface of $V$ with the surface normal $\vec n=\vec e_z$. The electric scalar potential $\varphi$ has to be known on the surface only. It is determined by the equations
\begin{alignat*}{2}
\nabla^2\varphi(\vec r) &= 
  \nabla\cdot\bigl(\vec v(\vec r)\times\vec B(\vec r)\bigr)
  &&\qquad \text{in $V$,} \\
\vec n(\vec s) \cdot \nabla\varphi(\vec s) &= 
  \vec n(\vec s) \cdot \bigl(\vec v(\vec s)\times\vec B(\vec s)\bigr)
  &&\qquad \text{on $S$,} 
\end{alignat*}
stemming from the conservation of charge $\nabla\cdot\vec j=0$ and the boundary condition of the eddy currents, $\vec n\cdot\vec j=0$. These equations can be solved using the Green's function of the three-dimensional Poisson equation \mbox{$G(\vec r, \vec r')=-1/(4\pi|\vec r-\vec r'|)$} \citep{Vladimirov1972, Stefani1999}:
\begin{multline}
\label{eq:LFVpen_scalPot}
\varphi(\vec s) =
  \int_V \frac{\bigl(\vec v(\vec r')\times\vec B(\vec r')\bigr)
                \cdot(\vec s-\vec r')}{2\pi|\vec s-\vec r'|^3} \de V' \\
  - \int_S \varphi(\vec s')\frac{\vec n(\vec s')\cdot(\vec s-\vec s')}
                                  {2\pi|\vec s-\vec s'|^3} \de S' \,.
\end{multline}
Since $\vec s,\vec s'\in S$ we have $\vec n\cdot(\vec s-\vec s')=0$ and the second term vanishes. We now rename $\vec s\to\vec s'$ and $\vec r' \to \vec r$ in~\eqref{eq:LFVpen_scalPot}, insert it into~\eqref{eq:LFVpen_force1} and swap the volume and surface integrals of the second term
\begin{multline*}
\vec F_L = \int_V v_x(z) \biggl[
  \sigma \bigl(\vec e_x\times\vec B(\vec r)\bigr)\times\vec B(\vec r) \\
  - \sigma\int_S 
    \frac{\bigl(\vec e_x\times\vec B(\vec r)\bigr)\cdot(\vec s'-\vec r)}
         {2\pi|\vec s'-\vec r|^3} 
  \bigl(\vec e_z\times\vec B(\vec s')\bigr) \de S' \biggr] \de V \,.
\end{multline*}
The integrand has the form $v_x(z)\vec w(\vec r)$: The velocity profile is weighed by a sensitivity function $\vec w$ (all terms within the square brackets) that is independent of the flow profile $v_x(z)$ and dependent on the geometry, the magnetic field and the flow direction. Since $v_x$ is independent of $x$ and $y$, the respective parts of the volume integration only apply to $\vec w$:
\begin{align*}
\vec F_L &= \int_{-\infty}^0 v_x(z)\tilde{\vec w}(z) \de z \,, &
\tilde{\vec w}(z) &= 
  \int_{-\infty}^\infty \int_{-\infty}^\infty \vec w(\vec r) \de x\de y \,.
\end{align*}
We now specify the permanent magnet as a cubic magnet with side length $2l$ and magnetization $\vec M=M\vec e_z$ parallel to one of its sides. Its centre is at $\vec r_M=(0,0,h)$, where $h>l$. The magnetic field in the fluid is \citep{Furlani2001}
\begin{align}
\notag
\vec B(\vec r) &= 
  -\frac{\mu_0M}{4\pi} \hat{\vec B}(\vec r-\vec r_M, \vec r')
  \Big|_{x'=-l}^l \Big|_{y'=-l}^l \Big|_{z'=-l}^{l} \,, \\[.5em]
\notag
\hat{\vec B}(\vec r, \vec r') &=
\begin{pmatrix}
    \artanh\left(\frac{y-y'}{|\vec r-\vec r'|}\right) \\
    \artanh\left(\frac{x-x'}{|\vec r-\vec r'|}\right) \\
    - \arctan\left(\frac{(x-x')(y-y')}{(z-z')|\vec r-\vec r'|}\right)
  \end{pmatrix} \,.
\end{align}
It exhibits the following symmetries: $B_{y/z}(x,y,z) = B_{y/z}(-x,y,z)$ and $B_x(x,y,z) = -B_x(-x,y,z)$. With these symmetries it can be shown, that $\tilde w_y = \tilde w_z = 0$ since the integrands are antisymmetric in $x$ and/or $x'$, so that the integrals over $x$ and $x'$ vanish. This leaves only a force component $F_{L,x}$ in flow direction with the weight-function
\begin{multline}
\notag
\tilde w_x(z) = -\sigma\int_{-\infty}^\infty \int_{-\infty}^\infty
  \biggl[B_y(\vec r)^2 + B_z(\vec r)^2 \\
  + \int_{-\infty}^\infty \int_{-\infty}^\infty
  \frac{B_z(\vec r)(y'-y)}{2\pi|\vec s'-\vec r|^3}
  B_y(\vec s') \de x' \de y' \biggr]_{z'=0} \de x \de y \,.
\end{multline}
Here, we also used that $B_y(x,y,z) = -B_y(x,-y,z)$ to eliminate another term in the surface integral. This formula applies for all magnetic fields that have the same symmetries as listed above (e.g. for a magnetic dipole in $z$-direction). These integrals have to be evaluated numerically. Here, they are calculated using the trapezoidal rule on grids for $x$, $y$, $x'$ and $y'$ that cluster near the magnet position $x_M=0$ and $y_M=0$. 121 points per integral were distributed over a domain of $\pm 70$\,mm for every integration. The result is displayed in Figure~\ref{fig:LFVpen_result}.\par
The strongest contribution of the flow to $\vec F_L$ is near the surface and the sensitivity rapidly decreases with increasing depth. $\tilde w_x$ is always negative, which is not immediately apparent from the surface integral. This means the Lorentz force opposes the flow, as was expected. To quantify a penetration depth of the LLFV we calculate the cumulative relative contribution to the final signal with increasing depth
\begin{align}
\tilde P(z) = \frac{\int_z^0 \tilde w_x(z') \de z'}
                   {\int_{-\infty}^0 \tilde w_x(z') \de z'} \,, 
  \qquad z \le 0 \,.
\end{align}
We see in Figure~\ref{fig:LFVpen_result} that $90\,\%$ of the LLFV-signal comes from the fluid layer with a thickness of 5.7\,mm below the top wall. This value is used as a length scale for calculating the interaction parameter $N$ in section~\ref{sec:lfvudv}.
\bibliography{}

\begin{thebibliography}{40}%
\makeatletter
\providecommand \@ifxundefined [1]{%
 \@ifx{#1\undefined}
}%
\providecommand \@ifnum [1]{%
 \ifnum #1\expandafter \@firstoftwo
 \else \expandafter \@secondoftwo
 \fi
}%
\providecommand \@ifx [1]{%
 \ifx #1\expandafter \@firstoftwo
 \else \expandafter \@secondoftwo
 \fi
}%
\providecommand \natexlab [1]{#1}%
\providecommand \enquote  [1]{``#1''}%
\providecommand \bibnamefont  [1]{#1}%
\providecommand \bibfnamefont [1]{#1}%
\providecommand \citenamefont [1]{#1}%
\providecommand \href@noop [0]{\@secondoftwo}%
\providecommand \href [0]{\begingroup \@sanitize@url \@href}%
\providecommand \@href[1]{\@@startlink{#1}\@@href}%
\providecommand \@@href[1]{\endgroup#1\@@endlink}%
\providecommand \@sanitize@url [0]{\catcode `\\12\catcode `\$12\catcode
  `\&12\catcode `\#12\catcode `\^12\catcode `\_12\catcode `\%12\relax}%
\providecommand \@@startlink[1]{}%
\providecommand \@@endlink[0]{}%
\providecommand \url  [0]{\begingroup\@sanitize@url \@url }%
\providecommand \@url [1]{\endgroup\@href {#1}{\urlprefix }}%
\providecommand \urlprefix  [0]{URL }%
\providecommand \Eprint [0]{\href }%
\providecommand \doibase [0]{http://dx.doi.org/}%
\providecommand \selectlanguage [0]{\@gobble}%
\providecommand \bibinfo  [0]{\@secondoftwo}%
\providecommand \bibfield  [0]{\@secondoftwo}%
\providecommand \translation [1]{[#1]}%
\providecommand \BibitemOpen [0]{}%
\providecommand \bibitemStop [0]{}%
\providecommand \bibitemNoStop [0]{.\EOS\space}%
\providecommand \EOS [0]{\spacefactor3000\relax}%
\providecommand \BibitemShut  [1]{\csname bibitem#1\endcsname}%
\let\auto@bib@innerbib\@empty
\bibitem [{\citenamefont {Davidson}(2001)}]{Davidson2001}%
  \BibitemOpen
  \bibfield  {author} {\bibinfo {author} {\bibfnamefont {P.~A.}\ \bibnamefont
  {Davidson}},\ }\href@noop {} {{\selectlanguage {english}\emph {\bibinfo
  {title} {An {{Introduction}} to {{Magnetohydrodynamics}}}}}},\ \bibinfo
  {edition} {1st}\ ed.,\ \bibinfo {series} {Cambridge Texts in Applied
  Mathematics}, Vol.~\bibinfo {volume} {25}\ (\bibinfo  {publisher} {{Cambridge
  University Press}},\ \bibinfo {address} {Cambridge, United Kingdom},\
  \bibinfo {year} {2001})\BibitemShut {NoStop}%
\bibitem [{\citenamefont {Asai}(2012)}]{Asai2012}%
  \BibitemOpen
  \bibfield  {author} {\bibinfo {author} {\bibfnamefont {S.}~\bibnamefont
  {Asai}},\ }\href@noop {} {{\selectlanguage {english}\emph {\bibinfo {title}
  {Electromagnetic {{Processing}} of {{Materials}}}}}},\ \bibinfo {series}
  {Fluid Mechanics and Its Applications}, Vol.~\bibinfo {volume} {99}\
  (\bibinfo  {publisher} {{Springer Netherlands}},\ \bibinfo {address}
  {Dordrecht},\ \bibinfo {year} {2012})\BibitemShut {NoStop}%
\bibitem [{\citenamefont {Shevchenko}\ \emph {et~al.}(2013)\citenamefont
  {Shevchenko}, \citenamefont {Boden}, \citenamefont {Eckert}, \citenamefont
  {Borin}, \citenamefont {Heinze},\ and\ \citenamefont
  {Odenbach}}]{Shevchenko2013}%
  \BibitemOpen
  \bibfield  {author} {\bibinfo {author} {\bibfnamefont {N.}~\bibnamefont
  {Shevchenko}}, \bibinfo {author} {\bibfnamefont {S.}~\bibnamefont {Boden}},
  \bibinfo {author} {\bibfnamefont {S.}~\bibnamefont {Eckert}}, \bibinfo
  {author} {\bibfnamefont {D.}~\bibnamefont {Borin}}, \bibinfo {author}
  {\bibfnamefont {M.}~\bibnamefont {Heinze}}, \ and\ \bibinfo {author}
  {\bibfnamefont {S.}~\bibnamefont {Odenbach}},\ }\href@noop {} {\bibfield
  {journal} {\bibinfo  {journal} {Eur. Phys. J. Spec. Top.}\ }\textbf {\bibinfo
  {volume} {220}},\ \bibinfo {pages} {63} (\bibinfo {year} {2013})}\BibitemShut
  {NoStop}%
\bibitem [{\citenamefont {Kelley}\ and\ \citenamefont
  {Sadoway}(2014)}]{Kelley2014}%
  \BibitemOpen
  \bibfield  {author} {\bibinfo {author} {\bibfnamefont {D.~H.}\ \bibnamefont
  {Kelley}}\ and\ \bibinfo {author} {\bibfnamefont {D.~R.}\ \bibnamefont
  {Sadoway}},\ }\href@noop {} {\bibfield  {journal} {\bibinfo  {journal} {Phys.
  Fluids}\ }\textbf {\bibinfo {volume} {26}},\ \bibinfo {pages} {057102}
  (\bibinfo {year} {2014})}\BibitemShut {NoStop}%
\bibitem [{\citenamefont {Chill{\`a}}\ and\ \citenamefont
  {Schumacher}(2012)}]{Chilla2012}%
  \BibitemOpen
  \bibfield  {author} {\bibinfo {author} {\bibfnamefont {F.}~\bibnamefont
  {Chill{\`a}}}\ and\ \bibinfo {author} {\bibfnamefont {J.}~\bibnamefont
  {Schumacher}},\ }\href@noop {} {\bibfield  {journal} {\bibinfo  {journal}
  {Eur. Phys. J. E.}\ }\textbf {\bibinfo {volume} {35}},\ \bibinfo {pages} {58}
  (\bibinfo {year} {2012})}\BibitemShut {NoStop}%
\bibitem [{\citenamefont {Adrian}\ and\ \citenamefont
  {Westerweel}(2011)}]{Adrian2011}%
  \BibitemOpen
  \bibfield  {author} {\bibinfo {author} {\bibfnamefont {R.~J.}\ \bibnamefont
  {Adrian}}\ and\ \bibinfo {author} {\bibfnamefont {J.}~\bibnamefont
  {Westerweel}},\ }\href@noop {} {{\selectlanguage {english}\emph {\bibinfo
  {title} {Particle {{Image Velocimetry}}}}}},\ Cambridge Aerospace Series\
  (\bibinfo  {publisher} {{Cambridge University Press}},\ \bibinfo {year}
  {2011})\BibitemShut {NoStop}%
\bibitem [{\citenamefont {Takeda}(1986)}]{Takeda1986}%
  \BibitemOpen
  \bibfield  {author} {\bibinfo {author} {\bibfnamefont {Y.}~\bibnamefont
  {Takeda}},\ }\href@noop {} {\bibfield  {journal} {\bibinfo  {journal} {Int.
  J. Heat Fluid Fl.}\ }\textbf {\bibinfo {volume} {7}},\ \bibinfo {pages} {313}
  (\bibinfo {year} {1986})}\BibitemShut {NoStop}%
\bibitem [{\citenamefont {Brito}\ \emph {et~al.}(2001)\citenamefont {Brito},
  \citenamefont {Nataf}, \citenamefont {Cardin}, \citenamefont {Aubert},\ and\
  \citenamefont {Masson}}]{Brito2001}%
  \BibitemOpen
  \bibfield  {author} {\bibinfo {author} {\bibfnamefont {D.}~\bibnamefont
  {Brito}}, \bibinfo {author} {\bibfnamefont {H.-C.}\ \bibnamefont {Nataf}},
  \bibinfo {author} {\bibfnamefont {P.}~\bibnamefont {Cardin}}, \bibinfo
  {author} {\bibfnamefont {J.}~\bibnamefont {Aubert}}, \ and\ \bibinfo {author}
  {\bibfnamefont {J.-P.}\ \bibnamefont {Masson}},\ }\href@noop {} {\bibfield
  {journal} {\bibinfo  {journal} {Exp. Fluids}\ }\textbf {\bibinfo {volume}
  {31}},\ \bibinfo {pages} {653} (\bibinfo {year} {2001})}\BibitemShut
  {NoStop}%
\bibitem [{\citenamefont {Eckert}\ and\ \citenamefont
  {Gerbeth}(2002)}]{Eckert2002}%
  \BibitemOpen
  \bibfield  {author} {\bibinfo {author} {\bibfnamefont {S.}~\bibnamefont
  {Eckert}}\ and\ \bibinfo {author} {\bibfnamefont {G.}~\bibnamefont
  {Gerbeth}},\ }\href@noop {} {\bibfield  {journal} {\bibinfo  {journal} {Exp.
  Fluids}\ }\textbf {\bibinfo {volume} {32}},\ \bibinfo {pages} {542} (\bibinfo
  {year} {2002})}\BibitemShut {NoStop}%
\bibitem [{\citenamefont {Boden}\ \emph {et~al.}(2008)\citenamefont {Boden},
  \citenamefont {Eckert}, \citenamefont {Willers},\ and\ \citenamefont
  {Gerbeth}}]{Boden2008}%
  \BibitemOpen
  \bibfield  {author} {\bibinfo {author} {\bibfnamefont {S.}~\bibnamefont
  {Boden}}, \bibinfo {author} {\bibfnamefont {S.}~\bibnamefont {Eckert}},
  \bibinfo {author} {\bibfnamefont {B.}~\bibnamefont {Willers}}, \ and\
  \bibinfo {author} {\bibfnamefont {G.}~\bibnamefont {Gerbeth}},\ }\href@noop
  {} {\bibfield  {journal} {\bibinfo  {journal} {Metallurgical and Materials
  Transactions A}\ }\textbf {\bibinfo {volume} {39}},\ \bibinfo {pages} {613}
  (\bibinfo {year} {2008})}\BibitemShut {NoStop}%
\bibitem [{\citenamefont {Ricou}\ and\ \citenamefont
  {Vives}(1982)}]{Ricou1982}%
  \BibitemOpen
  \bibfield  {author} {\bibinfo {author} {\bibfnamefont {R.}~\bibnamefont
  {Ricou}}\ and\ \bibinfo {author} {\bibfnamefont {C.}~\bibnamefont {Vives}},\
  }\href@noop {} {\bibfield  {journal} {\bibinfo  {journal} {Int. J. Heat Mass
  Transfer}\ }\textbf {\bibinfo {volume} {25}},\ \bibinfo {pages} {1579}
  (\bibinfo {year} {1982})}\BibitemShut {NoStop}%
\bibitem [{\citenamefont {Baker}\ \emph {et~al.}(2017)\citenamefont {Baker},
  \citenamefont {Poth{\'e}rat}, \citenamefont {Davoust}, \citenamefont
  {Debray},\ and\ \citenamefont {Klein}}]{Baker2017}%
  \BibitemOpen
  \bibfield  {author} {\bibinfo {author} {\bibfnamefont {N.~T.}\ \bibnamefont
  {Baker}}, \bibinfo {author} {\bibfnamefont {A.}~\bibnamefont {Poth{\'e}rat}},
  \bibinfo {author} {\bibfnamefont {L.}~\bibnamefont {Davoust}}, \bibinfo
  {author} {\bibfnamefont {F.}~\bibnamefont {Debray}}, \ and\ \bibinfo {author}
  {\bibfnamefont {R.}~\bibnamefont {Klein}},\ }\href@noop {} {\bibfield
  {journal} {\bibinfo  {journal} {Exp. Fluids}\ }\textbf {\bibinfo {volume}
  {58}} (\bibinfo {year} {2017})}\BibitemShut {NoStop}%
\bibitem [{\citenamefont {Wondrak}\ \emph {et~al.}(2017)\citenamefont
  {Wondrak}, \citenamefont {Pal}, \citenamefont {Stefani}, \citenamefont
  {Galindo},\ and\ \citenamefont {Eckert}}]{Wondrak2017}%
  \BibitemOpen
  \bibfield  {author} {\bibinfo {author} {\bibfnamefont {T.}~\bibnamefont
  {Wondrak}}, \bibinfo {author} {\bibfnamefont {J.}~\bibnamefont {Pal}},
  \bibinfo {author} {\bibfnamefont {F.}~\bibnamefont {Stefani}}, \bibinfo
  {author} {\bibfnamefont {V.}~\bibnamefont {Galindo}}, \ and\ \bibinfo
  {author} {\bibfnamefont {S.}~\bibnamefont {Eckert}},\ }\href@noop {}
  {\bibfield  {journal} {\bibinfo  {journal} {Flow Meas. Instrum.}\ } (\bibinfo
  {year} {2017})}\BibitemShut {NoStop}%
\bibitem [{\citenamefont {Heinicke}(2013)}]{Heinicke2013a}%
  \BibitemOpen
  \bibfield  {author} {\bibinfo {author} {\bibfnamefont {C.}~\bibnamefont
  {Heinicke}},\ }\href@noop {} {\bibfield  {journal} {\bibinfo  {journal} {Exp.
  Fluids}\ }\textbf {\bibinfo {volume} {54}},\ \bibinfo {pages} {1} (\bibinfo
  {year} {2013})}\BibitemShut {NoStop}%
\bibitem [{\citenamefont {Thess}\ \emph {et~al.}(2006)\citenamefont {Thess},
  \citenamefont {Votyakov},\ and\ \citenamefont {Kolesnikov}}]{Thess2006}%
  \BibitemOpen
  \bibfield  {author} {\bibinfo {author} {\bibfnamefont {A.}~\bibnamefont
  {Thess}}, \bibinfo {author} {\bibfnamefont {E.~V.}\ \bibnamefont {Votyakov}},
  \ and\ \bibinfo {author} {\bibfnamefont {Y.}~\bibnamefont {Kolesnikov}},\
  }\href@noop {} {\bibfield  {journal} {\bibinfo  {journal} {Phys. Rev. Lett.}\
  }\textbf {\bibinfo {volume} {96}} (\bibinfo {year} {2006})}\BibitemShut
  {NoStop}%
\bibitem [{\citenamefont {Thess}\ \emph {et~al.}(2007)\citenamefont {Thess},
  \citenamefont {Votyakov}, \citenamefont {Knaepen},\ and\ \citenamefont
  {Zikanov}}]{Thess2007}%
  \BibitemOpen
  \bibfield  {author} {\bibinfo {author} {\bibfnamefont {A.}~\bibnamefont
  {Thess}}, \bibinfo {author} {\bibfnamefont {E.~V.}\ \bibnamefont {Votyakov}},
  \bibinfo {author} {\bibfnamefont {B.}~\bibnamefont {Knaepen}}, \ and\
  \bibinfo {author} {\bibfnamefont {O.}~\bibnamefont {Zikanov}},\ }\href@noop
  {} {\bibfield  {journal} {\bibinfo  {journal} {New J. Phys.}\ }\textbf
  {\bibinfo {volume} {9}},\ \bibinfo {pages} {299} (\bibinfo {year}
  {2007})}\BibitemShut {NoStop}%
\bibitem [{\citenamefont {Heinicke}\ \emph {et~al.}(2012)\citenamefont
  {Heinicke}, \citenamefont {Tympel}, \citenamefont {Pulugundla}, \citenamefont
  {Rahneberg}, \citenamefont {Boeck},\ and\ \citenamefont
  {Thess}}]{Heinicke2012}%
  \BibitemOpen
  \bibfield  {author} {\bibinfo {author} {\bibfnamefont {C.}~\bibnamefont
  {Heinicke}}, \bibinfo {author} {\bibfnamefont {S.}~\bibnamefont {Tympel}},
  \bibinfo {author} {\bibfnamefont {G.}~\bibnamefont {Pulugundla}}, \bibinfo
  {author} {\bibfnamefont {I.}~\bibnamefont {Rahneberg}}, \bibinfo {author}
  {\bibfnamefont {T.}~\bibnamefont {Boeck}}, \ and\ \bibinfo {author}
  {\bibfnamefont {A.}~\bibnamefont {Thess}},\ }\href@noop {} {\bibfield
  {journal} {\bibinfo  {journal} {J. Appl. Phys.}\ }\textbf {\bibinfo {volume}
  {112}},\ \bibinfo {pages} {124914} (\bibinfo {year} {2012})}\BibitemShut
  {NoStop}%
\bibitem [{\citenamefont {Wang}\ \emph {et~al.}(2011)\citenamefont {Wang},
  \citenamefont {Klein}, \citenamefont {Kolesnikov},\ and\ \citenamefont
  {Thess}}]{Wang2011a}%
  \BibitemOpen
  \bibfield  {author} {\bibinfo {author} {\bibfnamefont {X.~D.}\ \bibnamefont
  {Wang}}, \bibinfo {author} {\bibfnamefont {R.}~\bibnamefont {Klein}},
  \bibinfo {author} {\bibfnamefont {Y.}~\bibnamefont {Kolesnikov}}, \ and\
  \bibinfo {author} {\bibfnamefont {A.}~\bibnamefont {Thess}},\ }\href@noop {}
  {\bibfield  {journal} {\bibinfo  {journal} {Mater. Sci. Forum}\ }\textbf
  {\bibinfo {volume} {690}},\ \bibinfo {pages} {99} (\bibinfo {year}
  {2011})}\BibitemShut {NoStop}%
\bibitem [{\citenamefont {Wegfrass}\ \emph {et~al.}(2012)\citenamefont
  {Wegfrass}, \citenamefont {Diethold}, \citenamefont {Werner}, \citenamefont
  {Fr{\"o}hlich}, \citenamefont {Halbedel}, \citenamefont {Hilbrunner},
  \citenamefont {Resagk},\ and\ \citenamefont {Thess}}]{Wegfrass2012}%
  \BibitemOpen
  \bibfield  {author} {\bibinfo {author} {\bibfnamefont {A.}~\bibnamefont
  {Wegfrass}}, \bibinfo {author} {\bibfnamefont {C.}~\bibnamefont {Diethold}},
  \bibinfo {author} {\bibfnamefont {M.}~\bibnamefont {Werner}}, \bibinfo
  {author} {\bibfnamefont {T.}~\bibnamefont {Fr{\"o}hlich}}, \bibinfo {author}
  {\bibfnamefont {B.}~\bibnamefont {Halbedel}}, \bibinfo {author}
  {\bibfnamefont {F.}~\bibnamefont {Hilbrunner}}, \bibinfo {author}
  {\bibfnamefont {C.}~\bibnamefont {Resagk}}, \ and\ \bibinfo {author}
  {\bibfnamefont {A.}~\bibnamefont {Thess}},\ }\href@noop {} {\bibfield
  {journal} {\bibinfo  {journal} {Appl. Phys. Lett.}\ }\textbf {\bibinfo
  {volume} {100}},\ \bibinfo {pages} {194103} (\bibinfo {year}
  {2012})}\BibitemShut {NoStop}%
\bibitem [{\citenamefont {Vasilyan}\ and\ \citenamefont
  {Fr{\"o}hlich}(2014)}]{Vasilyan2014}%
  \BibitemOpen
  \bibfield  {author} {\bibinfo {author} {\bibfnamefont {S.}~\bibnamefont
  {Vasilyan}}\ and\ \bibinfo {author} {\bibfnamefont {T.}~\bibnamefont
  {Fr{\"o}hlich}},\ }\href@noop {} {\bibfield  {journal} {\bibinfo  {journal}
  {Appl. Phys. Lett.}\ }\textbf {\bibinfo {volume} {105}},\ \bibinfo {pages}
  {223510} (\bibinfo {year} {2014})}\BibitemShut {NoStop}%
\bibitem [{\citenamefont {Wiederhold}\ \emph {et~al.}(2016)\citenamefont
  {Wiederhold}, \citenamefont {Ebert}, \citenamefont {Weidner}, \citenamefont
  {Halbedel}, \citenamefont {Fr{\"o}hlich},\ and\ \citenamefont
  {Resagk}}]{Wiederhold2016a}%
  \BibitemOpen
  \bibfield  {author} {\bibinfo {author} {\bibfnamefont {A.}~\bibnamefont
  {Wiederhold}}, \bibinfo {author} {\bibfnamefont {R.}~\bibnamefont {Ebert}},
  \bibinfo {author} {\bibfnamefont {M.}~\bibnamefont {Weidner}}, \bibinfo
  {author} {\bibfnamefont {B.}~\bibnamefont {Halbedel}}, \bibinfo {author}
  {\bibfnamefont {T.}~\bibnamefont {Fr{\"o}hlich}}, \ and\ \bibinfo {author}
  {\bibfnamefont {C.}~\bibnamefont {Resagk}},\ }\href@noop {} {\bibfield
  {journal} {\bibinfo  {journal} {Meas. Sci. Technol.}\ }\textbf {\bibinfo
  {volume} {27}},\ \bibinfo {pages} {125306} (\bibinfo {year}
  {2016})}\BibitemShut {NoStop}%
\bibitem [{\citenamefont {Sokolov}\ \emph {et~al.}(2016)\citenamefont
  {Sokolov}, \citenamefont {Noskov}, \citenamefont {Pavlinov},\ and\
  \citenamefont {Kolesnikov}}]{Sokolov2016}%
  \BibitemOpen
  \bibfield  {author} {\bibinfo {author} {\bibfnamefont {I.}~\bibnamefont
  {Sokolov}}, \bibinfo {author} {\bibfnamefont {V.}~\bibnamefont {Noskov}},
  \bibinfo {author} {\bibfnamefont {A.}~\bibnamefont {Pavlinov}}, \ and\
  \bibinfo {author} {\bibfnamefont {Y.}~\bibnamefont {Kolesnikov}},\
  }\href@noop {} {\bibfield  {journal} {\bibinfo  {journal}
  {Magnetohydrodynamics}\ }\textbf {\bibinfo {volume} {52}},\ \bibinfo {pages}
  {481} (\bibinfo {year} {2016})}\BibitemShut {NoStop}%
\bibitem [{\citenamefont {Hern{\'a}ndez}\ \emph {et~al.}(2016)\citenamefont
  {Hern{\'a}ndez}, \citenamefont {Schleichert}, \citenamefont {Karcher},
  \citenamefont {Fr{\"o}hlich}, \citenamefont {Wondrak},\ and\ \citenamefont
  {Timmel}}]{Hernandez2016a}%
  \BibitemOpen
  \bibfield  {author} {\bibinfo {author} {\bibfnamefont {D.}~\bibnamefont
  {Hern{\'a}ndez}}, \bibinfo {author} {\bibfnamefont {J.}~\bibnamefont
  {Schleichert}}, \bibinfo {author} {\bibfnamefont {C.}~\bibnamefont
  {Karcher}}, \bibinfo {author} {\bibfnamefont {T.}~\bibnamefont
  {Fr{\"o}hlich}}, \bibinfo {author} {\bibfnamefont {T.}~\bibnamefont
  {Wondrak}}, \ and\ \bibinfo {author} {\bibfnamefont {K.}~\bibnamefont
  {Timmel}},\ }\href@noop {} {\bibfield  {journal} {\bibinfo  {journal} {Meas.
  Sci. Technol.}\ }\textbf {\bibinfo {volume} {27}},\ \bibinfo {pages} {065302}
  (\bibinfo {year} {2016})}\BibitemShut {NoStop}%
\bibitem [{\citenamefont {Ng}\ \emph {et~al.}(2013)\citenamefont {Ng},
  \citenamefont {Chung},\ and\ \citenamefont {Ooi}}]{Ng2013}%
  \BibitemOpen
  \bibfield  {author} {\bibinfo {author} {\bibfnamefont {C.~S.}\ \bibnamefont
  {Ng}}, \bibinfo {author} {\bibfnamefont {D.}~\bibnamefont {Chung}}, \ and\
  \bibinfo {author} {\bibfnamefont {A.}~\bibnamefont {Ooi}},\ }\href@noop {}
  {\bibfield  {journal} {\bibinfo  {journal} {Int. J. Heat Fluid Fl.}\ }\textbf
  {\bibinfo {volume} {44}},\ \bibinfo {pages} {554} (\bibinfo {year}
  {2013})}\BibitemShut {NoStop}%
\bibitem [{\citenamefont {Ng}\ \emph {et~al.}(2015)\citenamefont {Ng},
  \citenamefont {Ooi}, \citenamefont {Lohse},\ and\ \citenamefont
  {Chung}}]{Ng2015}%
  \BibitemOpen
  \bibfield  {author} {\bibinfo {author} {\bibfnamefont {C.~S.}\ \bibnamefont
  {Ng}}, \bibinfo {author} {\bibfnamefont {A.}~\bibnamefont {Ooi}}, \bibinfo
  {author} {\bibfnamefont {D.}~\bibnamefont {Lohse}}, \ and\ \bibinfo {author}
  {\bibfnamefont {D.}~\bibnamefont {Chung}},\ }\href@noop {} {\bibfield
  {journal} {\bibinfo  {journal} {J. Fluid Mech.}\ }\textbf {\bibinfo {volume}
  {764}},\ \bibinfo {pages} {349} (\bibinfo {year} {2015})}\BibitemShut
  {NoStop}%
\bibitem [{\citenamefont {Shishkina}(2016)}]{Shishkina2016a}%
  \BibitemOpen
  \bibfield  {author} {\bibinfo {author} {\bibfnamefont {O.}~\bibnamefont
  {Shishkina}},\ }\href@noop {} {\bibfield  {journal} {\bibinfo  {journal}
  {Phys. Rev. E}\ }\textbf {\bibinfo {volume} {93}},\ \bibinfo {pages}
  {051102(R)} (\bibinfo {year} {2016})}\BibitemShut {NoStop}%
\bibitem [{\citenamefont {Brown}\ and\ \citenamefont
  {Ahlers}(2006)}]{Brown2006}%
  \BibitemOpen
  \bibfield  {author} {\bibinfo {author} {\bibfnamefont {E.}~\bibnamefont
  {Brown}}\ and\ \bibinfo {author} {\bibfnamefont {G.}~\bibnamefont {Ahlers}},\
  }\href@noop {} {\bibfield  {journal} {\bibinfo  {journal} {J. Fluid Mech.}\
  }\textbf {\bibinfo {volume} {568}},\ \bibinfo {pages} {351} (\bibinfo {year}
  {2006})}\BibitemShut {NoStop}%
\bibitem [{\citenamefont {Zhou}\ \emph {et~al.}(2009)\citenamefont {Zhou},
  \citenamefont {Xi}, \citenamefont {Zhou}, \citenamefont {Sun},\ and\
  \citenamefont {Xia}}]{Zhou2009}%
  \BibitemOpen
  \bibfield  {author} {\bibinfo {author} {\bibfnamefont {Q.}~\bibnamefont
  {Zhou}}, \bibinfo {author} {\bibfnamefont {H.-D.}\ \bibnamefont {Xi}},
  \bibinfo {author} {\bibfnamefont {S.-Q.}\ \bibnamefont {Zhou}}, \bibinfo
  {author} {\bibfnamefont {C.}~\bibnamefont {Sun}}, \ and\ \bibinfo {author}
  {\bibfnamefont {K.-Q.}\ \bibnamefont {Xia}},\ }\href@noop {} {\bibfield
  {journal} {\bibinfo  {journal} {J. Fluid Mech.}\ }\textbf {\bibinfo {volume}
  {630}},\ \bibinfo {pages} {367} (\bibinfo {year} {2009})}\BibitemShut
  {NoStop}%
\bibitem [{\citenamefont {Vogt}\ \emph {et~al.}(2013)\citenamefont {Vogt},
  \citenamefont {Grants}, \citenamefont {Eckert},\ and\ \citenamefont
  {Gerbeth}}]{Vogt2013}%
  \BibitemOpen
  \bibfield  {author} {\bibinfo {author} {\bibfnamefont {T.}~\bibnamefont
  {Vogt}}, \bibinfo {author} {\bibfnamefont {I.}~\bibnamefont {Grants}},
  \bibinfo {author} {\bibfnamefont {S.}~\bibnamefont {Eckert}}, \ and\ \bibinfo
  {author} {\bibfnamefont {G.}~\bibnamefont {Gerbeth}},\ }\href@noop {}
  {\bibfield  {journal} {\bibinfo  {journal} {J. Fluid Mech.}\ }\textbf
  {\bibinfo {volume} {736}},\ \bibinfo {pages} {641} (\bibinfo {year}
  {2013})}\BibitemShut {NoStop}%
\bibitem [{\citenamefont {Vogt}\ \emph {et~al.}(2014)\citenamefont {Vogt},
  \citenamefont {R{\"a}biger},\ and\ \citenamefont {Eckert}}]{Vogt2014a}%
  \BibitemOpen
  \bibfield  {author} {\bibinfo {author} {\bibfnamefont {T.}~\bibnamefont
  {Vogt}}, \bibinfo {author} {\bibfnamefont {D.}~\bibnamefont {R{\"a}biger}}, \
  and\ \bibinfo {author} {\bibfnamefont {S.}~\bibnamefont {Eckert}},\
  }\href@noop {} {\bibfield  {journal} {\bibinfo  {journal} {J. Fluid Mech.}\
  }\textbf {\bibinfo {volume} {753}},\ \bibinfo {pages} {472} (\bibinfo {year}
  {2014})}\BibitemShut {NoStop}%
\bibitem [{\citenamefont {Tasaka}\ \emph {et~al.}(2016)\citenamefont {Tasaka},
  \citenamefont {Igaki}, \citenamefont {Yanagisawa}, \citenamefont {Vogt},
  \citenamefont {Z{\"u}rner},\ and\ \citenamefont {Eckert}}]{Tasaka2016}%
  \BibitemOpen
  \bibfield  {author} {\bibinfo {author} {\bibfnamefont {Y.}~\bibnamefont
  {Tasaka}}, \bibinfo {author} {\bibfnamefont {K.}~\bibnamefont {Igaki}},
  \bibinfo {author} {\bibfnamefont {T.}~\bibnamefont {Yanagisawa}}, \bibinfo
  {author} {\bibfnamefont {T.}~\bibnamefont {Vogt}}, \bibinfo {author}
  {\bibfnamefont {T.}~\bibnamefont {Z{\"u}rner}}, \ and\ \bibinfo {author}
  {\bibfnamefont {S.}~\bibnamefont {Eckert}},\ }\href@noop {} {\bibfield
  {journal} {\bibinfo  {journal} {Phys. Rev. E}\ }\textbf {\bibinfo {volume}
  {93}},\ \bibinfo {pages} {043109} (\bibinfo {year} {2016})}\BibitemShut
  {NoStop}%
\bibitem [{\citenamefont {Plevachuk}\ \emph {et~al.}(2014)\citenamefont
  {Plevachuk}, \citenamefont {Sklyarchuk}, \citenamefont {Eckert},
  \citenamefont {Gerbeth},\ and\ \citenamefont {Novakovic}}]{Plevachuk2014}%
  \BibitemOpen
  \bibfield  {author} {\bibinfo {author} {\bibfnamefont {Y.}~\bibnamefont
  {Plevachuk}}, \bibinfo {author} {\bibfnamefont {V.}~\bibnamefont
  {Sklyarchuk}}, \bibinfo {author} {\bibfnamefont {S.}~\bibnamefont {Eckert}},
  \bibinfo {author} {\bibfnamefont {G.}~\bibnamefont {Gerbeth}}, \ and\
  \bibinfo {author} {\bibfnamefont {R.}~\bibnamefont {Novakovic}},\ }\href@noop
  {} {\bibfield  {journal} {\bibinfo  {journal} {J. Chem. Eng. Data}\ }\textbf
  {\bibinfo {volume} {59}},\ \bibinfo {pages} {757} (\bibinfo {year}
  {2014})}\BibitemShut {NoStop}%
\bibitem [{\citenamefont {{\c C}engel}(2008)}]{Cengel2008}%
  \BibitemOpen
  \bibfield  {author} {\bibinfo {author} {\bibfnamefont {Y.~A.}\ \bibnamefont
  {{\c C}engel}},\ }\href@noop {} {{\selectlanguage {english}\emph {\bibinfo
  {title} {Introduction to {{Thermodynamics}} and {{Heat Transfer}}}}}},\
  \bibinfo {edition} {2nd}\ ed.\ (\bibinfo  {publisher} {{McGraw-Hill
  Primis}},\ \bibinfo {year} {2008})\BibitemShut {NoStop}%
\bibitem [{\citenamefont {Batchelor}(1954)}]{Batchelor1954}%
  \BibitemOpen
  \bibfield  {author} {\bibinfo {author} {\bibfnamefont {G.~K.}\ \bibnamefont
  {Batchelor}},\ }\href@noop {} {\bibfield  {journal} {\bibinfo  {journal} {Q.
  Appl. Math.}\ }\textbf {\bibinfo {volume} {12}},\ \bibinfo {pages} {209}
  (\bibinfo {year} {1954})}\BibitemShut {NoStop}%
\bibitem [{\citenamefont {Yu}\ \emph {et~al.}(2007)\citenamefont {Yu},
  \citenamefont {Li},\ and\ \citenamefont {Ecke}}]{Yu2007}%
  \BibitemOpen
  \bibfield  {author} {\bibinfo {author} {\bibfnamefont {H.}~\bibnamefont
  {Yu}}, \bibinfo {author} {\bibfnamefont {N.}~\bibnamefont {Li}}, \ and\
  \bibinfo {author} {\bibfnamefont {R.~E.}\ \bibnamefont {Ecke}},\ }\href@noop
  {} {\bibfield  {journal} {\bibinfo  {journal} {Phys. Rev. E}\ }\textbf
  {\bibinfo {volume} {76}},\ \bibinfo {pages} {026303} (\bibinfo {year}
  {2007})}\BibitemShut {NoStop}%
\bibitem [{\citenamefont {Scheel}\ and\ \citenamefont
  {Schumacher}(2016)}]{Scheel2016}%
  \BibitemOpen
  \bibfield  {author} {\bibinfo {author} {\bibfnamefont {J.~D.}\ \bibnamefont
  {Scheel}}\ and\ \bibinfo {author} {\bibfnamefont {J.}~\bibnamefont
  {Schumacher}},\ }\href@noop {} {\bibfield  {journal} {\bibinfo  {journal} {J.
  Fluid Mech.}\ }\textbf {\bibinfo {volume} {802}},\ \bibinfo {pages} {147}
  (\bibinfo {year} {2016})}\BibitemShut {NoStop}%
\bibitem [{\citenamefont {Dubovikova}\ \emph {et~al.}(2016)\citenamefont
  {Dubovikova}, \citenamefont {Resagk}, \citenamefont {Karcher},\ and\
  \citenamefont {Kolesnikov}}]{Dubovikova2016}%
  \BibitemOpen
  \bibfield  {author} {\bibinfo {author} {\bibfnamefont {N.}~\bibnamefont
  {Dubovikova}}, \bibinfo {author} {\bibfnamefont {C.}~\bibnamefont {Resagk}},
  \bibinfo {author} {\bibfnamefont {C.}~\bibnamefont {Karcher}}, \ and\
  \bibinfo {author} {\bibfnamefont {Y.}~\bibnamefont {Kolesnikov}},\
  }\href@noop {} {\bibfield  {journal} {\bibinfo  {journal} {Meas. Sci.
  Technol.}\ }\textbf {\bibinfo {volume} {27}},\ \bibinfo {pages} {055102}
  (\bibinfo {year} {2016})}\BibitemShut {NoStop}%
\bibitem [{\citenamefont {Vladimirov}(1972)}]{Vladimirov1972}%
  \BibitemOpen
  \bibfield  {author} {\bibinfo {author} {\bibfnamefont {V.~S.}\ \bibnamefont
  {Vladimirov}},\ }\href@noop {} {{\selectlanguage {german}\emph {\bibinfo
  {title} {{Gleichungen der mathematischen Physik}}}}},\ \bibinfo {series}
  {Hochschulb{\"u}cher f{\"u}r Mathematik}, Vol.~\bibinfo {volume} {74}\
  (\bibinfo  {publisher} {{VEB Deutscher Verlag der Wissenschaften}},\ \bibinfo
  {address} {Berlin},\ \bibinfo {year} {1972})\BibitemShut {NoStop}%
\bibitem [{\citenamefont {Stefani}\ and\ \citenamefont
  {Gerbeth}(1999)}]{Stefani1999}%
  \BibitemOpen
  \bibfield  {author} {\bibinfo {author} {\bibfnamefont {F.}~\bibnamefont
  {Stefani}}\ and\ \bibinfo {author} {\bibfnamefont {G.}~\bibnamefont
  {Gerbeth}},\ }\href@noop {} {\bibfield  {journal} {\bibinfo  {journal}
  {Inverse Probl.}\ }\textbf {\bibinfo {volume} {15}},\ \bibinfo {pages} {771}
  (\bibinfo {year} {1999})}\BibitemShut {NoStop}%
\bibitem [{\citenamefont {Furlani}(2001)}]{Furlani2001}%
  \BibitemOpen
  \bibfield  {author} {\bibinfo {author} {\bibfnamefont {E.~P.}\ \bibnamefont
  {Furlani}},\ }\href@noop {} {{\selectlanguage {english}\emph {\bibinfo
  {title} {Permanent {{Magnet}} and {{Electromechanical Devices}}:
  {{Materials}}, {{Analysis}}, and {{Applications}}}}}},\ Electromagnetism\
  (\bibinfo  {publisher} {{Academic Press, Inc.}},\ \bibinfo {address} {San
  Diego},\ \bibinfo {year} {2001})\BibitemShut {NoStop}%
\end{thebibliography}%
\end{document}